\documentclass[aps,preprintnumbers,amsmath,amssymb,nofootinbib,superscriptaddress,notitlepage,11pt]{revtex4-1}
\usepackage{epsfig}
\usepackage[utf8]{inputenc}
\usepackage{subfigure}
\usepackage{multirow}
\usepackage{float}
\usepackage[figuresright]{rotating}
\usepackage{graphicx}
\usepackage{color}
\usepackage{xcolor}
\usepackage[normalem]{ulem}
\usepackage[
    colorlinks=true,
    linkcolor=blue,
    breaklinks=true,
    urlcolor=magenta,
    citecolor=blue]{hyperref}
\usepackage{pifont}
\usepackage{booktabs}
\usepackage{multirow}
\usepackage{longtable,booktabs}
\usepackage{diagbox}
\usepackage{afterpage}

\newcommand{\itp}{\affiliation{CAS Key Laboratory of Theoretical Physics, Institute of Theoretical Physics,\\ Chinese Academy of Sciences, Beijing 100190, China}}

\newcommand{\ucas}{\affiliation{School of Physical Sciences, University of Chinese Academy of Sciences, Beijing 100049, China}}

\newcommand{\qfnu}{\affiliation{College of Physics and Engineering, Qufu Normal University, Qufu 273165, China}}

\newcommand{\peng}{\affiliation{Peng Huanwu Collaborative Center for Research and Education, Beihang University, Beijing 100191, China}}

\begin{document}

\title{Radiative decays of the heavy-quark-spin molecular partner of $T_{cc}^+$} 

\author{Zhao-Sai Jia}\email{jzsqfphys@163.com}\qfnu \itp
\author{Zhen-Hua Zhang}\email{zhangzhenhua@itp.ac.cn}\itp \ucas
\author{Gang Li}\email{gli@qfnu.edu.cn} \qfnu \itp
\author{Feng-Kun Guo} \email{fkguo@itp.ac.cn}
\itp \ucas \peng

\date{\today}

\begin{abstract} 
  \rule{0ex}{3ex}
With the assumptions that the $T_{cc}^+$ discovered at LHCb is a $D^{*}D$ hadronic molecule, using a nonrelativistic effective field theory we calculate the radiative partial widths of $T_{cc}^* \to D^*D\gamma$ with $T_{cc}^*$ being a $D^{*}D^{*}$ shallow bound state and the heavy-quark-spin partner of $T_{cc}^+$. The $I=0$ $D^*D$ rescattering effect with the $T_{cc}$ pole is taken into account. 
The results show that the isoscalar $D^{\ast} D$ rescattering can increase the tree-level decay width of $T_{cc}^{\ast +}\rightarrow D^{*+}D^0\gamma$ by about $50\%$ and decrease that of $T_{cc}^{\ast +}\rightarrow D^{*0}D^+\gamma$ by a similar amount. The two-body partial decay widths of the $T_{cc}^{*+}$ into $T_{cc}^+\gamma$ and $T_{cc}^+\pi^0$ are also calculated, and the results are about $6$ and $3~\rm{keV}$, respectively. Considering that the $D^*$ needs to be reconstructed from the $D\pi$ or $D\gamma$ final state in an experimental measurement, the four-body partial widths of the $T_{cc}^{*+}$ into $DD\gamma\gamma$ and $DD\pi\gamma$ are explicitly calculated, and we find that the interference effect between different intermediate $D^*D\gamma$ states is small. The total radiative decay width of the $T_{cc}^*$ is predicted to be about $24~\rm{keV}$. Adding the hadronic decay widths of $T_{cc}^* \to D^*D\pi$, the total width of the $T_{cc}^*$ is finally predicted to be  $(65\pm2)$~keV. 
\end{abstract}

\maketitle

\section{Introduction}
The LHCb Collaboration has reported a narrow resonance, the double-charm exotic candidate $T_{cc}$ with probable quantum numbers $I(J^P)=0(1^+)$, in the $D^0D^0\pi^+$ invariant mass distribution~\cite{LHCb:2021vvq,LHCb:2021auc}. Its mass and decay width were reported as~\cite{LHCb:2021vvq,LHCb:2021auc}
\begin{align}
    \delta m_{\rm BW}&= m_{\rm BW}-(m_{D^{*+}}+m_{D^0}) =-273 \pm 61 \pm 5_{-14}^{+11}\, \rm{keV},\nonumber\\
    \Gamma_{\rm BW}&=410\pm 165\pm43_{-38}^{+18}\, \rm{keV},
    \label{Eq:BW1}
\end{align}
parametrizing the $T_{cc}^+$ using a relativistic $P$-wave two-body Breit-Wigner function with a Blatt-Weisskopf form factor, and
\begin{align}
    \delta m_{\rm pole}&=m_{\rm pole}-(m_{D^{*+}}+m_{D^0}) =-360\pm 40_{-0}^{+4}\, \rm{keV},\nonumber\\
    \Gamma_{\rm pole}&=48\pm 2_{-14}^{+~0}\, \rm{keV},
    \label{Eq:UBW2}
\end{align}
using a unitarized Breit-Wigner profile~\cite{LHCb:2021auc}.
An analysis of the LHCb data with the full $DD\pi$ three-body effects taken into account gives~\cite{Du:2021zzh} 
\begin{align}
    \delta m_{\rm pole}=-356_{-38}^{+39}~\rm{keV}, \quad \Gamma_{\rm pole} = (56 \pm 2)~\rm{keV}. \label{eq:Tcc_3b}
\end{align}
By analyzing the line shape of the $T_{cc}$ or the low-energy $S$-wave $DD^*$ scattering parameters~\cite{Albaladejo:2021vln,Feijoo:2021ppq,Du:2021zzh,Dai:2023cyo,Wang:2023ovj}, it has been concluded that the $T_{cc}$ is an excellent candidate of a $DD^*$ hadronic molecule~\cite{Meng:2021jnw,Ling:2021bir,Dong:2021bvy,Ren:2021dsi,Xin:2021wcr,Chen:2022asf}. It was predicted to have a heavy-quark-spin symmetry (HQSS)
partner  $T_{cc}^*$, a $D^*D^*$ hadronic molecule with the quantum numbers $I(J^P)=0(1^+)$~\cite{Du:2021zzh, Albaladejo:2021vln}.
In particular, the mass of the $T_{cc}^{\ast}$ relative to the $D^{\ast}D^{\ast}$ threshold is predicted to be $\mathcal{B} = 2m_{D^*} - m_{T_{cc}^*}=(503 \pm 40) \, \rm{keV}$ in Ref.~\cite{Du:2021zzh}, which is called the binding energy of the $T_{cc}^{\ast}$ in the following. 
Precise knowledge of the $T_{cc}^*$ decay width is valuable for its searching in experiments, and it can be calculated in a nonrelativistic effective field theory called XEFT.

The XEFT is a nonrelativistic effective field theory which was first constructed to systematically study the long-range properties of the exotic $X(3872)$~\cite{Belle:2003nnu,ParticleDataGroup:2022pth}, also known as $\chi_{c1}(3872)$ with a mass coinciding with the $D^0\bar{D}^{\ast 0}$ threshold.
The $D$, $D^{\ast}$, $\bar{D}$, $\bar{D}^{\ast }$, and pions are the effective degrees of freedom in XEFT and are all treated nonrelativistically~\cite{Fleming:2007rp}. 
The partial decay widths of the $T_{cc}$, including $T_{cc} \to DD\pi$ and $T_{cc} \to DD\gamma$, are calculated using XEFT in Ref.~\cite{Fleming:2021wmk}; the result of the total width of $T_{cc}$ about 58~keV is in good agreement with Eq.~\eqref{eq:Tcc_3b} extracted in Ref.~\cite{Du:2021zzh}. In Ref.~\cite{Dai:2023mxm}, the next-to-leading-order (NLO) contributions to the strong decay width of the $T_{cc}$ are calculated including the contributions from one-pion exchange and final-state interaction (FSI). 
In Ref.~\cite{Jia:2022qwr}, we have 
calculated the hadronic partial widths of the spin partner $T_{cc}^{\ast}$ decaying into $D^{\ast}D\pi(\to DD\pi\pi)$ including contributions of the $D^{\ast} D$ and $D^{\ast} \pi$ FSIs. 
The total hadronic decay width of the $T_{cc}^{*}$ is predicted to be about $41~\rm{keV}$. 
The $T_{cc}^*$ can also decay radiatively into the $D^*D\gamma$ (subsequently to $DD\pi\gamma$ and $DD\gamma\gamma$)
final states. In this work, we compute the partial widths of such radiative decays and will give the total width of the $T_{cc}^*$ by summing up the hadronic and radiative partial widths.

This paper is organized as follows. In Sec.~\ref{sec:LAGRANGIAN}, we introduce the XEFT effective Lagrangian for the charmed mesons, photons, and pions and the power counting of the Feynman diagrams for the $T_{cc}^{\ast}\to D^{\ast}D\gamma$ processes. The amplitudes and partial decay rates of the $T_{cc}^{\ast} \to D^{\ast} D \gamma$ including contributions from the $D^{\ast} D$ FSIs  are derived in Sec.~\ref{sec:AMPLITUDES3B}. The amplitudes and partial decay rates of the $T_{cc}^* \to T_{cc}\gamma$ and $T_{cc}^* \to T_{cc}\pi$ are derived in Sec.~\ref{sec:AMPLITUDES2B}, and the numerical results for the partial decay widths of the $T_{cc}^{*+}$ are shown in Sec.~\ref{sec:Results3B2B}. The four-body decays $T_{cc}^{*} \to DD\gamma\gamma$ and $DD\gamma\pi$ including the corrections from the $D^*D$ and $D^*\pi$ FSIs are discussed in Sec.~\ref{sec:Results4B}. Finally, all the results are summarized in Sec.~\ref{sec:SUMMARY}. Some expressions are relegated to the Appendixes.

\section{Effective Lagrangian and Power Counting}\label{sec:LAGRANGIAN}

In this section, the effective Lagrangian and the power counting rules of the diagrams for the decays of the $T_{cc}^{\ast+}$ are introduced. For the $T_{cc}^{*+}$ being an $S$-wave isoscalar $D^*D^*$ shallow bound state with quantum numbers $J^P=1^+$ and a binding energy $\mathcal{B}=(503 \pm 40) ~ \rm{keV}$~\cite{Du:2021zzh}, the typical momentum and velocity of the $D^*$ mesons in $T_{cc}^{*+}$ are  
$p_{D^*}\simeq \gamma_{D^*D^*}\equiv \sqrt{2\mu_{D^*D^*}\mathcal{B}}\lesssim 33~\rm{MeV}$ and $v_{D^*}\simeq\sqrt{{\mathcal{B}/}({2\mu_{D^*D^*})}}\lesssim 0.02$, respectively, where $\mu_{D^*D^*}$ is the reduced mass of $D^{*+}$ and $D^{*0}$. Therefore, the $D^*$ and $D$ mesons can be treated nonrelativistically in the decays of $T_{cc}^* \to D^*D\gamma$, $DD\gamma\gamma$, and $DD\gamma\pi$. 
The maximal energy of the emitted pion in the $T_{cc}^* \to DD\gamma\pi$ decays is 
\begin{align}
    E_{\pi}=\frac{(m_{T_{cc}^*}-2m_{D})^2+m_{\pi}^2}{2(m_{T_{cc}^{*}}-2m_D)} \simeq 177~\rm{MeV},
\end{align}
where $m_{T_{cc}^*}, m_{D}, m_{D^*}$, and $m_{\pi}$ are the masses of $T_{cc}^*$, $D$, $D^*$, and $\pi$, respectively.
Then the momentum of the emitted pion in the rest frame of the $T_{cc}^*$ is $p_{\pi}\lesssim 110~\rm{MeV}$. 
Despite that, the pion in triangle loops (involving the $D^*\pi$ rescattering) will still be treated nonrelativistically, while the phase spaces of the decays are treated relativistically. 
Since such diagrams provide only a small correction (to be calculated later), this simplification presents a good approximation. 

The XEFT Lagrangian we use for the decays of $T_{cc}^{*}$ reads~\cite{Fleming:2007rp,Guo:2017jvc}
\begin{align}
\mathcal{L}=&\, H^{\ast i\, \dagger}\left(i \partial^{0}+\frac{\nabla^{2}}{2 m_{H^{\ast}}}\right) H^{\ast i}+H^{\dagger}\left(i \partial^{0}+\frac{\nabla^{2}}{2 m_{H}}\right) H +\frac{1}{2}\left\langle\pi^{\dagger}\left(i \partial^{0}+\frac{\nabla^{2}}{2 m_{\pi}}+\delta\right) \pi \right\rangle\nonumber\\
&-C_{0}\left(H^{\ast i\, T} \tau_{2} H^{\ast i}\right)^{\dagger}\left(H^{\ast j\,T} \tau_{2} H^{\ast j}\right)-C_{1}\left(H^{\ast i\, T} \tau_{2} \tau_{a} H^{\ast i}\right)^{\dagger}\left(H^{\ast j\,T} \tau_{2} \tau_{a} H^{\ast j}\right)\nonumber\\
&+\frac{\bar{g}}{F_{\pi} \sqrt{2m_{\pi}}} \left(H^{\dagger} \partial^{i} \pi H^{\ast i}+\mathrm{H.c.} \right)+\frac{1}{2} {H}^{\dagger}\mu_{D}{\mathcal{B}}^{i}{H}^{*i}+\frac{C_{0D}}{2}(H^T\tau_2H^{\ast i})^ {\dagger}(H^T\tau_2 H^{\ast i})\nonumber\\
&+\frac{C_{\frac{1}{2}\pi}}{6m_{\pi}}\left(\pi\tau_1H^{*i}\right)^\dagger\left(\pi\tau_1H^{*i}\right)+\frac{C_{\frac{3}{2}\pi}}{12m_{\pi}}\left\{\left[\left(\pi\tau_3+\frac{1}{2}\langle\pi\tau_3\rangle\right)\tau_1 H^{*i}\right]^{\dagger}\left[\left(\pi\tau_3+\frac{1}{2}\langle\pi\tau_3\rangle\right)\tau_1 H^{*i}\right]\right.\nonumber\\
&\left.+3\left[\left(\pi\tau_3-\frac{1}{2}\langle\pi\tau_3\rangle\right) H^{*i}\right]^{\dagger}\left[\left(\pi\tau_3-\frac{1}{2}\langle\pi\tau_3\rangle\right) H^{*i}\right]\right\},
\label{eq:XEFTlagrangian}
\end{align}
with the pseudoscalar $H=(D^0, D^+)^T$, vector heavy mesons $H^*=(D^{*0},D^{*+})^T$, the magnetic field $B^k=\epsilon^{ijk}\partial^iA^j$, and the pions
\begin{align}
\pi=\left(\begin{array}{cc}
\pi^{0} & \sqrt{2}\pi^{+} \\
\sqrt{2}\pi^{-} & -\pi^{0}
\end{array}\right).
\end{align}
Here $m_{H}$, $m_{H^{\ast}}$, and $m_{\pi}$ are the masses of the $H$, $H^{\ast}$, and $\pi$ particles, respectively; 
$\delta=\Delta-m_\pi\simeq 7~\rm{MeV}$ with $\Delta=m_{D^{*0}}-m_{D^0}$ comes from a field redefinition shifting the residual $D^*$ mass from the $D^*$ kinetic term to the pion kinetic term~\cite{Fleming:2007rp}, and it introduces a small momentum scale $\mu=\sqrt{\Delta^2-m_{\pi}^2}\simeq\sqrt{2m_{\pi}\delta}\approx 45~\rm{MeV}$ appearing in the pion propagator~\cite{Fleming:2007rp,Dai:2019hrf}; the pion decay constant is taken as $F_{\pi}=92.2~\rm{MeV}$, and $\tau_a$ with $a=1,2,3$ are the Pauli matrices in the isospin space in which the traces $\left(\langle\, \rangle\right)$ act.

\begin{figure}
    \centering
    \includegraphics[scale=0.34]{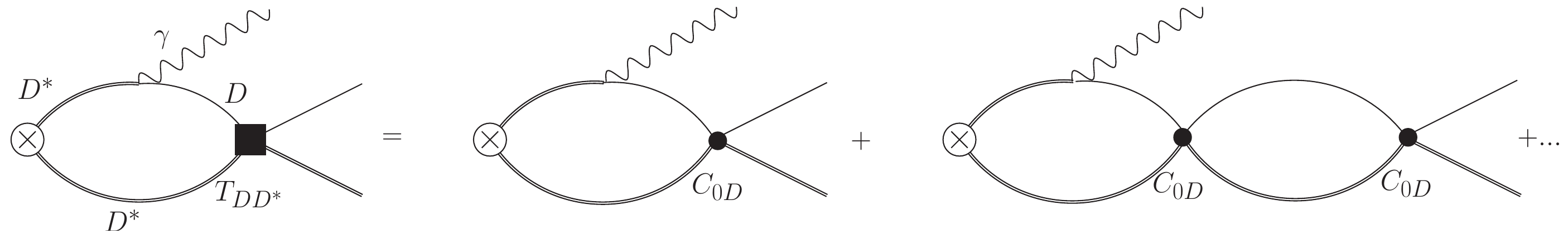}
    \caption{Resumming the $D^*D$ rescattering diagrams. The single thin lines represent the $D^+(D^0)$, the double lines represent the $D^{*0}(D^{*+})$, and the wavy lines represent the photon.}
    \label{Fig.Resummation}
\end{figure}

The first line in Eq.~\eqref{eq:XEFTlagrangian} includes the kinetic terms for the charmed mesons and pions. The second line contains the contact interactions of the $D^{*+}$ and $D^{*0}$, where the first term mediates the $D^{\ast}D^{\ast}$ scattering in the isoscalar channel and the second term mediates the scattering in the isovector channel. In the third line, the first term describes the coupling between the charmed mesons and a pion, with the coupling constant $\bar{g}\simeq 0.27$;\footnote{Here, the $\bar{g}$ parameter is related to the $g$ parameter in Ref.~\cite{Fleming:2021wmk} by $\bar g=g/2$.} 
the second term gives the magnetic couplings for the charmed mesons and a photon~\cite{Amundson:1992yp,Stewart:1998ke,Fleming:2021wmk} with the matrix of transition magnetic moments $\mu_D=\text{diag}(\mu_{D^0},\mu_{D^+})$, where $\mu_{D^+}=-0.15 \, \rm{GeV}^{-1}$ and $\mu_{D^0}=0.55 \, \rm{GeV}^{-1}$ are obtained by reproducing the partial widths $\Gamma[D^{*+}\to D^+\gamma]=1.33 \, \rm{keV}$~\cite{Karliner:2021qok} and $\Gamma[D^{*0} \to D^0\gamma]=19.52 \, \rm{keV}$~\cite{Guo:2019qcn}; 
the last term is the isoscalar contact interaction for $D^{*}D\to D^{*}D$. Because of the existence of the $T_{cc}^+$, the resummation effect shown in Fig.~\ref{Fig.Resummation} needs to be considered~\cite{Dai:2019hrf} 
by replacing $C_{0D}$ with the near-threshold $T$ matrix for the isoscalar $D^{*}D\to D^{*}D$~\cite{Kaplan:1998we}
\begin{align}
	C_{0D}\to T_{DD^*}= -\frac{2\pi}{\mu_{DD^*}} \frac{1}{1/a+ i p}, 
	\label{Eq:unitary TDDstar}
\end{align}
where $\mu_{DD^*}$ is the reduced mass of $D^*$ and $D$, $p=\vert \vec{p}_{D^*}-\vec{p}_D \vert/2$ is the relative momentum between $D^*$ and $D$ in the $D^*D$ center-of-mass (c.m.) frame, and the $D^*D$ scattering length $a$ is set to be $a=\left[-\left(6.72^{+0.36}_{-0.45}\right)-i\left(0.10^{+0.03}_{-0.03}\right)\right]\, \rm{fm}$ obtained in the analysis in Ref.~\cite{Du:2021zzh}. 
Here the isospin-breaking effect, which is a higher-order effect~\cite{Guo:2019fdo}, is neglected in the isoscalar $D^*D\to D^*D$ rescattering. 
We also ignore the isovector $DD^*$ FSI, which is much weaker than the isoscalar one, since there is no evidence for an isovector double-charm tetraquark near the $D^*D$ threshold. The last two terms with $C_{\frac{1}{2}\pi}$ and $C_{\frac{3}{2}\pi}$ in Eq.~\eqref{eq:XEFTlagrangian} are the $D^*\pi \to D^*\pi$ contact interactions for $I={1}/{2}$ and $I={3}/{2}$, respectively, and the coefficients $C_{\frac{1}{2}\pi}=25.2~\mathrm{GeV}^{-1}$ and $C_{\frac{3}{2}\pi}=-6.8~\mathrm{GeV}^{-1}$ are derived by matching to the $D^*\pi$ scattering lengths, which should be approximately equal to the $D\pi$ ones in Ref.~\cite{Liu:2012zya} (for detailed derivations, see Appendix~A in Ref.~\cite{Jia:2022qwr}) due to HQSS.

The effective Lagrangian for the $T_{cc}^*$ coupling to $D^*D^*$ in $S$ wave can be written as 
\begin{align}
	\mathcal{L}_s=& \frac{g_s}{\sqrt{2}} \varepsilon^{ijk} T_{cc}^{\ast +,i}  D^{\ast+,j} D^{\ast0,k},
	\label{Eq:TsL0}
\end{align}
where $\varepsilon^{ijk}$ is the three-dimensional antisymmetric Levi-Civita tensor. 
The effective coupling $g_s$ can be derived from the residue of the $D^{*+}D^{*0}\to D^{*+}D^{*0}$ scattering amplitude at the $T_{cc}^{*+}$ pole as~\cite{Weinberg:1965zz,Baru:2003qq,Guo:2017jvc}
\begin{align}
{g_s^2} = \frac{2\pi \gamma_{D^*D^*}}{\mu^2_{D^*D^*}}.
\end{align} 

\begin{figure}[tb]
    \subfigure[] {
        \includegraphics[scale=1.2]{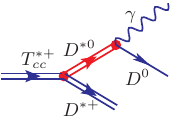}
        \label{fig_TsDscDgFDa}
    }
    \subfigure[] {
        \includegraphics[scale=1.2]{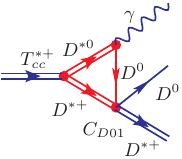}
        \label{fig_TsDscDgFDb}
    }
    \subfigure[] {
        \includegraphics[scale=1.2]{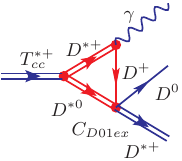}
        \label{fig_TsDscDgFDc}
    }\\
    \subfigure[] {
        \includegraphics[scale=1.2]{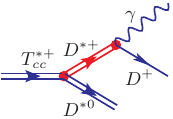}
        \label{fig_TsDsDcgFDa}
    }
    \subfigure[] {
        \includegraphics[scale=1.2]{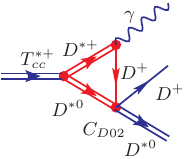}
        \label{fig_TsDsDcgFDb}
    }
    \subfigure[] {
        \includegraphics[scale=1.2]{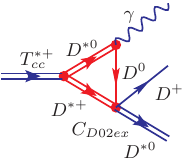}
        \label{fig_TsDsDcgFDc}
    }
    \caption{Feynman diagrams for calculating the partial decay width of $T_{cc}^{*+}\rightarrow D^*D\gamma$. The double lines represent the spin-1 mesons, $T_{cc}^{*+}$, $D^{*+}$, or $D^{*0}$; the single thin lines represent the pseudoscalar charmed mesons, $D^+$ or $D^0$; and the wavy lines represent the photon.}
    \label{fig_TsDsDgFD}
\end{figure}

With the above Lagrangians in Eqs.~\eqref{eq:XEFTlagrangian} and \eqref{Eq:TsL0}, the leading-order (LO) amplitude for the $T_{cc}^{*}\to D^*D\gamma$ including the effects of the $D^*D$ FSIs is shown in Fig.~\ref{fig_TsDsDgFD}, where Figs.~\ref{fig_TsDscDgFDa}-\ref{fig_TsDscDgFDc} are the diagrams for the decay $T_{cc}^{*+} \to D^{*+}D^0\gamma$ and Figs.~\ref{fig_TsDsDcgFDa}-\ref{fig_TsDsDcgFDc} are for the decay $T_{cc}^{*+} \to D^{*0}D^+\gamma$.

In the following, we will briefly introduce the power counting of all these diagrams in Fig.~\ref{fig_TsDsDgFD} following the analysis for the decays of the $X(3872)$ and $T_{cc}$ in Refs.~\cite{Fleming:2007rp,Dai:2019hrf,Yan:2021wdl}. The relevant small momenta involved in the decays of the $T_{cc}^*$ are $\{p_D,p_{D^{\ast}}, q_{\gamma}, \gamma, \mu\}$, where $q_{\gamma}$ is momentum of the emitted photon. They are of the same order and denoted generically by $Q$. 
Each nonrelativistic propagator is of $\mathcal{O}(Q^{-2})$, and, as the nonrelativistic energy is of $\mathcal{O}(Q^{2})$, each nonrelativistic loop integral measure counts as $\mathcal{O}(Q^{5})$. The isoscalar contact interaction $C_{0D}$ between the $D^{*}$ and $D$ is replaced with $T_{DD^*}$ in Eq.~\eqref{Eq:unitary TDDstar} and, thus, contributes at $\mathcal{O}(Q^{-1})$~\cite{Fleming:2007rp,Guo:2017jvc}.
For the diagrams in Fig.~\ref{fig_TsDsDgFD}, the amplitudes from  diagrams~\ref{fig_TsDscDgFDa} and \ref{fig_TsDsDcgFDa} scale as $\mathcal{O}(Q / Q^{2})=\mathcal{O}(Q^{-1})$, since there are one nonrelativistic propagator and one $P$-wave photon vertex. The amplitudes for diagrams~\ref{fig_TsDscDgFDb}, \ref{fig_TsDscDgFDc}, \ref{fig_TsDsDcgFDb}, and \ref{fig_TsDsDcgFDc} also scale as $\mathcal{O}( Q^{-1})$ for the decays $T_{cc}^{\ast+ }\to D^{*+}D^0\gamma$ and $T_{cc}^{\ast+ }\to D^{*0}D^+\gamma$ and contribute at LO.

\section{Differential decay rates of  $T_{cc}^{\ast}\to D^{\ast}D\gamma$}\label{sec:AMPLITUDES3B}

In this section, all the decay amplitudes of the $T_{cc}^{\ast+ }\to D^{*+}D^0\gamma$ and $T_{cc}^{\ast+ }\to D^{*0}D^+\gamma$ processes in Fig.~\ref{fig_TsDsDgFD} are given, as well as expressions of their partial differential decay rates. The Breit-Wigner form of the $D^*$ propagator, $G_{D^*}(p)$, is used to include the contribution of the $D^{*}$ self-energy, {\it i.e.},
\begin{align}
  G_{D^*}(p)=  \frac{i}{p^0_{D^*}-m_{D^*}-\frac{\vec{p}_{D^*}^{\, 2}}{2m_{D^*}}+i\frac{\Gamma_{D^*}}{2}},
  \label{Eq.Dstar_self_energy}
\end{align}
where $D^*$ denotes $D^{*+}$ or $D^{*0}$, $p=(p^0_{D^*},\vec{p}_{D^*})$ is the four-momentum of the $D^*$, $\Gamma_{D^{*+}}=83.4~\rm{keV}$~\cite{ParticleDataGroup:2022pth}, and $\Gamma_{D^{*0}}=55.3~\rm{keV}$~\cite{Guo:2019qcn}.

\subsection{Partial decay rate of $T_{cc}^{\ast +}\to D^{\ast +}D^0\gamma$}\label{sec:TsDscDgAmp}

First, we consider the three-body decay $T_{cc}^{\ast +}\to D^{\ast +}D^0\gamma$.
The LO amplitude from the tree diagram in Fig.~\ref{fig_TsDscDgFDa} reads
\begin{align}
&& i \mathcal{A}_{a}^{\text{3B}}=&\, \frac{-g_s \mu_{D^*D^*} \mu_{D^{0}}}{\sqrt{2}(\gamma_{D^*D^*}^2+\vec{p}_{D^{\ast+}}^{\, 2})}\left \{ \epsilon^{i}(T_{cc}^{\ast +})\epsilon^{j*}(D^{\ast +})q_{\gamma }^i\epsilon^{j*}(\gamma)-\epsilon^{i}(T_{cc}^{\ast +})\epsilon^{j*}(D^{\ast +})q_{\gamma}^j\epsilon^{i*}(\gamma) \right \},
\end{align}
where $\vec{p}_{D^{\ast +}}$ is the three-momentum of the external $D^{\ast+}$, $\vec{q}_{\gamma}$ is the three-momentum of the final state $\gamma$, and $\epsilon^i(T_{cc}^{*+})$,
$\epsilon^{j*}(D^{*+})$, and $\epsilon^{j(i)*}(\gamma)$ are the polarization vectors of the incoming particle $T_{cc}^{*+}$ and the outcoming particles $D^{*+}$ and $\gamma$, respectively.

The LO amplitudes from the $D^*D$ rescattering diagrams in Figs.~\ref{fig_TsDscDgFDb} and \ref{fig_TsDscDgFDc} read
\begin{align}
 i \mathcal{A}_{b}^{\text{3B}}=&\, \frac{-g_sC_{0D1}\mu_{D^0}}{2\sqrt{2}}I_{b}(q_{\gamma})\left \{ \epsilon^{i}(T_{cc}^{\ast +})\epsilon^{j*}(D^{\ast +})q_{\gamma}^i\epsilon^{j*}(\gamma)-\epsilon^{i}(T_{cc}^{\ast +})\epsilon^{j*}(D^{\ast +})q_{\gamma}^j\epsilon^{i*}(\gamma)\right \},\\
 i \mathcal{A}_{c}^{\text{3B}}=&\, \frac{g_sC_{0D1ex}\mu_{D^+}}{2\sqrt{2}}I_{c}(q_{\gamma})\left \{ \epsilon^{i}(T_{cc}^{\ast +})\epsilon^{j*}(D^{\ast +})q_{\gamma}^i\epsilon^{j*}(\gamma)-\epsilon^{i}(T_{cc}^{\ast +})\epsilon^{j*}(D^{\ast +})q_{\gamma}^j\epsilon^{i*}(\gamma)\right \},
\end{align}
where $C_{0D1}=+\frac{1}{2} C_{0D}$ and $C_{0D1\text{ex}}=-\frac{1}{2} C_{0D}$ are the contact terms for the $D^{*+}D^0\to D^{*+}D^0$ and the $D^{*0}D^+\to D^{*+}D^0$, respectively, and $I_b(q_\gamma)$ and $I_c(q_\gamma)$ are the three-point scalar loop integrals, whose explicit expressions can be obtained from $I(q)$ given in Appendix~\ref{sec:Triangal loop}~\cite{Guo:2010ak, Dai:2019hrf} as follows:
$m_1$, $m_2$ and $m_3$ in Eq.~\eqref{Eq:loop_integral} are taken to be the masses of $D^{\ast 0}$, $D^{\ast +}$ and $D^0$ for $I_{b}(q_{\gamma})$, and the masses of $D^{\ast +}$, $D^{\ast 0}$ and $D^+$ for $I_{c}(q_{\gamma})$, respectively.

The decay rate is given by
\begin{align}
d\Gamma=&2M 2E_1 2E_2 \frac{1}{2M}\frac{1}{2j+1} \sum_{\mathrm{spins}} \vert \mathcal{A}\vert^2 d\Phi_{3},
\label{dGamma1}
\end{align}
where the overall factor comes from the normalization of nonrelativistic particles, with $M$ being the mass of the $T_{cc}^{*}$, $E_1$ and $E_2$ being the energies of two nonrelativistic final-state particles in the $T_{cc}^{*}$ rest frame, respectively, $j=1$ is the spin of the $T_{cc}^{*}$, and there is a sum over the polarizations of spin-1 particles. Here the three-body phase space integration is given by
\begin{align}
\int d\Phi_{3}=&\int\frac{1}{32 \pi^3} \frac{1}{4 E_1 E_2} d\vert \vec{p}_1 \vert^2 d\vert \vec{p}_2 \vert^2,
\label{Eq.three_body_phase_space}
\end{align}
where $\vec{p}_1$ and $\vec{p}_2$ are the three-momenta for two of the final-state particles in the rest frame of the initial particle.

The LO partial differential rate for the $T_{cc}^{\ast +} \to D^{\ast +}D^0\gamma$ including corrections from the $D^*D$ rescattering reads
\begin{align}
	\frac{d \Gamma_{T_{cc}^{\ast +} \to D^{\ast +}D^0\gamma}}{d|\vec{p}_{D^0}|^2d|\vec{p}_{D^{\ast +}}|^2}=&\frac{1}{96 \pi^3}\sum_{\mathrm{spins}} \left \vert \mathcal{A}_{a}^{\text{3B}}+\mathcal{A}_{b}^{\text{3B}}+\mathcal{A}_{c}^{\text{3B}} \right \vert^2.
\end{align}

\subsection{Partial decay rate of $T_{cc}^{\ast +}\to D^{\ast 0}D^+\gamma$}\label{sec:TsDsDcgAmp}
For the decay $T_{cc}^{\ast +}\to D^{\ast 0}D^+\gamma$, the LO amplitude from the tree diagram in Fig.~\ref{fig_TsDsDcgFDa} reads
\begin{align}
&& i \mathcal{A}_{d}^{\text{3B}}=&\, \frac{g_s \mu_{D^*D^*} \mu_{D^{+}}}{\sqrt{2}(\gamma_{D^*D^*}^2+\vec{p}_{D^{\ast0}}^{\, 2})}\left \{ \epsilon^{i}(T_{cc}^{\ast +})\epsilon^{j*}(D^{\ast 0})q_{\gamma}^i\epsilon^{j*}(\gamma)-\epsilon^{i}(T_{cc}^{\ast +})\epsilon^{j*}(D^{\ast 0})q_{\gamma}^j\epsilon^{i*}(\gamma)\right \},
\end{align}
where $\vec{p}_{D^{\ast 0}}$ and  $\epsilon^j(D^{*0})$ are the three-momentum and  polarization vector of the external $D^{\ast 0}$, respectively. The LO amplitudes from the $D^*D$ rescattering diagrams in Figs.~\ref{fig_TsDsDcgFDb} and \ref{fig_TsDsDcgFDc} are
\begin{align}
&& i \mathcal{A}_{e}^{\text{3B}}=&\, \frac{g_sC_{0D2}\mu_{D^+}}{2\sqrt{2}}I_c(q_{\gamma})\left \{ \epsilon^{i}(T_{cc}^{\ast +})\epsilon^{j*}(D^{\ast 0})q_{\gamma}^i\epsilon^{j*}(\gamma)-\epsilon^{i}(T_{cc}^{\ast +})\epsilon^{j*}(D^{\ast 0})q_{\gamma}^j\epsilon^{i*}(\gamma)\right \},\\
&& i \mathcal{A}_{f}^{\text{3B}}=&\, \frac{-g_sC_{0D2ex}\mu_{D^0}}{2\sqrt{2}}I_b(q_{\gamma})\left \{ \epsilon^{i}(T_{cc}^{\ast +})\epsilon^{j*}(D^{\ast 0})q_{\gamma}^i\epsilon^{j*}(\gamma)-\epsilon^{i}(T_{cc}^{\ast +})\epsilon^{j*}(D^{\ast 0})q_{\gamma}^j\epsilon^{i*}(\gamma)\right \},
\end{align}
where $C_{0D2}=\frac{1}{2}C_{0D}$ and $C_{0D2\text{ex}}=-\frac{1}{2}C_{0D}$ are the contact terms for the $D^{*0}D^+\to D^{*0}D^+$ and the $D^{*+}D^0\to D^{*0}D^+$, respectively. 

The LO partial differential rate for the $T_{cc}^{\ast +}\to D^{\ast 0} D^+ \gamma$ including the corrections from the $D^*D$ rescattering is 
\begin{align}
	\frac{d \Gamma_{T_{cc}^{\ast +}\to D^{\ast 0}D^+\gamma}}{d|\vec{p}_{D^+}|^2d|\vec{p}_{D^{\ast 0}}|^2} =\frac{1}{96 \pi^3}\sum_{\mathrm{spins}} \left \vert \mathcal{A}_{d}^{\text{3B}}+\mathcal{A}_{e}^{\text{3B}}+\mathcal{A}_{f}^{\text{3B}} \right \vert^2.
\end{align}

\section{partial decay widths of $T_{cc}^{\ast +}\to T_{cc}^+\gamma$ and $T_{cc}^+\pi^0$}\label{sec:AMPLITUDES2B}

\begin{figure}[htbp]
    \subfigure[] {
        \includegraphics[scale=1.2]{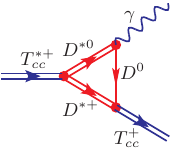}
        \label{fig_TsTgFDa}
    }
    \subfigure[] {
        \includegraphics[scale=1.2]{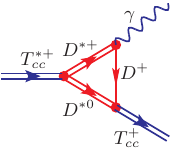}
        \label{fig_TsTgFDb}
    }
    \subfigure[] {
        \includegraphics[scale=1.2]{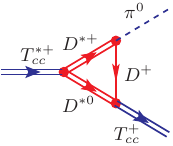}
        \label{fig_TsTpiFDc}
    }
    \subfigure[] {
        \includegraphics[scale=1.2]{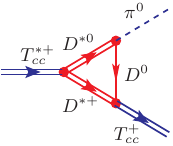}
        \label{fig_TsTpiFDd}
    }
    \caption{Feynman diagrams for calculating the partial decay widths of $T_{cc}^{*+}\rightarrow T_{cc}^{+}\gamma$ and $T_{cc}^{*+}\rightarrow T_{cc}^{+}\pi^0$. The double lines represent the spin-1 mesons, $T_{cc}^{*+}$, $T_{cc}^+$, $D^{*0}$ and $D^{*+}$; the single thin lines represent the pseudoscalar charmed mesons, $D^+$ and $D^0$; the wavy lines represent the photon; and the dashed lines represent the pion.}
    \label{Fig.TsTg/piFD}
\end{figure}

To show the $D^*D$ rescattering effects more clearly, in this section we consider the two-body decays $T_{cc}^{*+} \to T_{cc}^+ \gamma$ and $T_{cc}^{*+} \to T_{cc}^+ \pi^0$ shown in Fig.~\ref{Fig.TsTg/piFD}. The effective Lagrangian for the $T_{cc}^+$ coupling to $D^{*}D$ can be written as
\begin{align}
\mathcal{L}_0=&\, g_0 T_{cc}^{+ i}D^{*+i}D^{0}+ g_+ T_{cc}^{+ i}D^{*0i}D^{+},
\label{Eq:TL0}
\end{align}
where the coupling constants $g_0=1.03/\sqrt{2m_{T_{cc}^+}}$ and $g_+=-0.99/\sqrt{2m_{T_{cc}^+}}$ are taken from the analysis of scheme~\uppercase\expandafter{\romannumeral3} in Ref.~\cite{Du:2021zzh}.\footnote{We use nonrelativistic normalization for all the fields, and the mass dimension of $g_{0,+}$ is $-1/2$. The $1/{\sqrt{2m_{T_{cc}
^+}}}$ factor comes from the different normalizations of the fields used here and in Ref.~\cite{Du:2021zzh}. The imaginary parts of the couplings are neglected, which come from the $DD\pi$ three-body dynamics and are about 2 orders of magnitude smaller than the real parts. }

For the decay $T_{cc}^{\ast +}\to T_{cc}^+\gamma$, using the effective  Lagrangians in Eqs.~\eqref{eq:XEFTlagrangian} and ~\eqref{Eq:TL0}, the amplitudes in Figs.~\ref{fig_TsTgFDa} and~\ref{fig_TsTgFDb} are
\begin{align}
 && i\mathcal{A}_{a}^{\text{2B}}=&\, \frac{g_s g_+ \mu_{D^+}}{2\sqrt{2}} I_{b}(q_{\gamma}) \left \{ \epsilon^{i}(T_{cc}^{\ast +})\epsilon^{j*}(T_{cc}^+)q_{\gamma}^i\epsilon^{j*}(\gamma)-\epsilon^{i}(T_{cc}^{\ast +})\epsilon^{j*}(T_{cc}^+)q_{\gamma}^j\epsilon^{i*}(\gamma)\right \},\\
 && i\mathcal{A}_{b}^{\text{2B}}=&\, \frac{-g_s g_0 \mu_{D^0}}{2\sqrt{2}} I_{c}(q_{\gamma})\left \{\epsilon^{i}(T_{cc}^{\ast +})\epsilon^{j*}(T_{cc}^+)q_{\gamma}^i\epsilon^{j*}(\gamma)-\epsilon^{i}(T_{cc}^{\ast +})\epsilon^{j*}(T_{cc}^+)q_{\gamma}^j\epsilon^{i*}(\gamma)\right \},
\end{align}
where $\vec{q_{\gamma}}$ is the three-momentum of the external photon and the $\epsilon^{j*}(T_{cc}^+)$ is the polarization vector of the $T_{cc}^+$. 

The differential decay width is given by
\begin{align}
d\Gamma_{2B}=&2M 2E_1 2E_2 \frac{1}{2M}\frac{1}{2j+1} \sum_{\mathrm{spins}} \vert \mathcal{A}\vert^2 d\Phi_{2},
\label{dGamma1}
\end{align}
where the two-body phase space is
\begin{align}
\int d\Phi_{2}(P; p_1, p_2)=&\, \int d\Omega_1 \frac{\vert \vec{p}_1 \vert}{(2\pi)^2 4 E},
\label{Eq.two_body_phase_space}
\end{align}
where $|\vec{p}_1|$ is the magnitude of the three-momentum of particle 1 in the rest frame of the initial state, $d\Omega_1=d\cos\theta_1d\varphi_1$ is the solid angle of particle 1, and $E$ is the energy of the initial-state particle in the same reference frame.

The partial decay width for $T_{cc}^{\ast +} \to T_{cc}^+\gamma$ is 
\begin{align}
\Gamma_{T_{cc}^{\ast +}\to T_{cc}^+\gamma}
=&\, \frac{m_{T_{cc}^+} \vert \vec{q_{\gamma}} \vert}{6\pi m_{T_{cc}^{*+}}}\sum_{\mathrm{spins}} \left \vert \mathcal{A}_{a}^{\text{2B}} + \mathcal{A}_{b}^{\text{2B}} \right \vert^2 \, .
\end{align}

For the isospin-breaking decay $T_{cc}^{\ast +}\to T_{cc}^+\pi^0$, the amplitudes from the diagrams in Figs.~\ref{fig_TsTpiFDc} and~\ref{fig_TsTpiFDd} read
\begin{align}
 i\mathcal{A}_{c}^{\text{2B}}=&\, \frac{-g_s \bar{g} g_+}{2 F_{\pi} \sqrt{m_{\pi^0}} } I_{c}(p_{\pi^0})\epsilon^{ijk}\epsilon^i(T_{cc}^{\ast +})\epsilon^{j*}(T_{cc}^+)p_{\pi^0}^k,\\
 i\mathcal{A}_{d}^{\text{2B}}=&\,\frac{-g_s \bar{g} g_0}{2 F_{\pi} \sqrt{m_{\pi^0}} } I_{b}(p_{\pi^0})\epsilon^{ijk}\epsilon^i(T_{cc}^{\ast +})\epsilon^{j*}(T_{cc}^+)p_{\pi^0}^k,
\end{align}
where $\vec{p}_{\pi^0}$ is the three-momentum of the external $\pi^0$. The partial width of $T_{cc}^{\ast +} \to T_{cc}^+\pi^0$ is 
\begin{align}
\Gamma_{T_{cc}^{\ast +}\to T_{cc}^+\pi^0}
=&\, \frac{m_{\pi^0} m_{T_{cc}^+} \vert \vec{p}_{\pi^0}\vert}{3\pi m_{T_{cc}^{*+}}}\sum_{\mathrm{spins}} \left \vert \mathcal{A}_{c}^{\text{2B}}+\mathcal{A}_{d}^{\text{2B}} \right \vert^2 \, .
\end{align}
This decay breaks isospin symmetry; thus, if we use the isospin-averaged masses for all the involved mesons, the contributions in Figs.~\ref{fig_TsTpiFDc} and~\ref{fig_TsTpiFDd} would vanish.

\section{Partial decay widths for $T_{cc}^{\ast} \to D^{\ast}D\gamma$, $T_{cc}\gamma$, and $T_{cc}\pi$}\label{sec:Results3B2B}

\begin{table}[htbp]
\caption{\label{Tab:3B2BDW} Partial decay widths of the $T_{cc}^{\ast +}$ with a binding energy $\mathcal{B}=(503 \pm 40)\, \rm{keV}$. $\Gamma_{\text{Tree}}$ contains the contributions from the tree-level diagrams, and $\Gamma_{\text{LO}}$ is the LO decay width which includes the contributions from the tree-level and $D^{\ast}D$ rescattering diagrams. The errors come from that of the binding energy $\mathcal{B}$ predicted in Ref.~\cite{Du:2021zzh}.}
\renewcommand{\arraystretch}{1.2}
\setlength{\tabcolsep}{45pt}{
\begin{tabular*}{\columnwidth}{l|c|c}
\hline\hline  
        $\Gamma \, [\rm{keV}]$
        & $\Gamma_{\text{Tree}}$
        & $\Gamma_{\text{LO}}$
        \\[3pt]        
\hline
  $\Gamma[T_{cc}^{\ast +}\rightarrow D^{\ast +}D^0\gamma]$ & $15.6 \pm 0.2$ & $23.4 \pm 0.1$ \\[3pt]
\hline
  $\Gamma[T_{cc}^{\ast +}\rightarrow D^{\ast 0}D^+\gamma]$ & $1.0 \pm 0.1$ & $0.4 \pm 0.0$ \\[3pt]    
\hline
  $\Gamma[T_{cc}^{\ast +}\rightarrow T_{cc}^{+}\gamma]$ & ... &{$6.1 \pm 0.1$} \\[3pt]
\hline
  $\Gamma[T_{cc}^{\ast +}\rightarrow T_{cc}^{+}\pi^0]$ & ... &{$2.7 \pm 0.1$} \\[3pt]
\hline\hline
\end{tabular*}}
\end{table}

\begin{figure}[htbp]
    \subfigure[$T_{cc}^{\ast +}\to D^{\ast +}D^0\gamma$] {\includegraphics[scale=0.335]{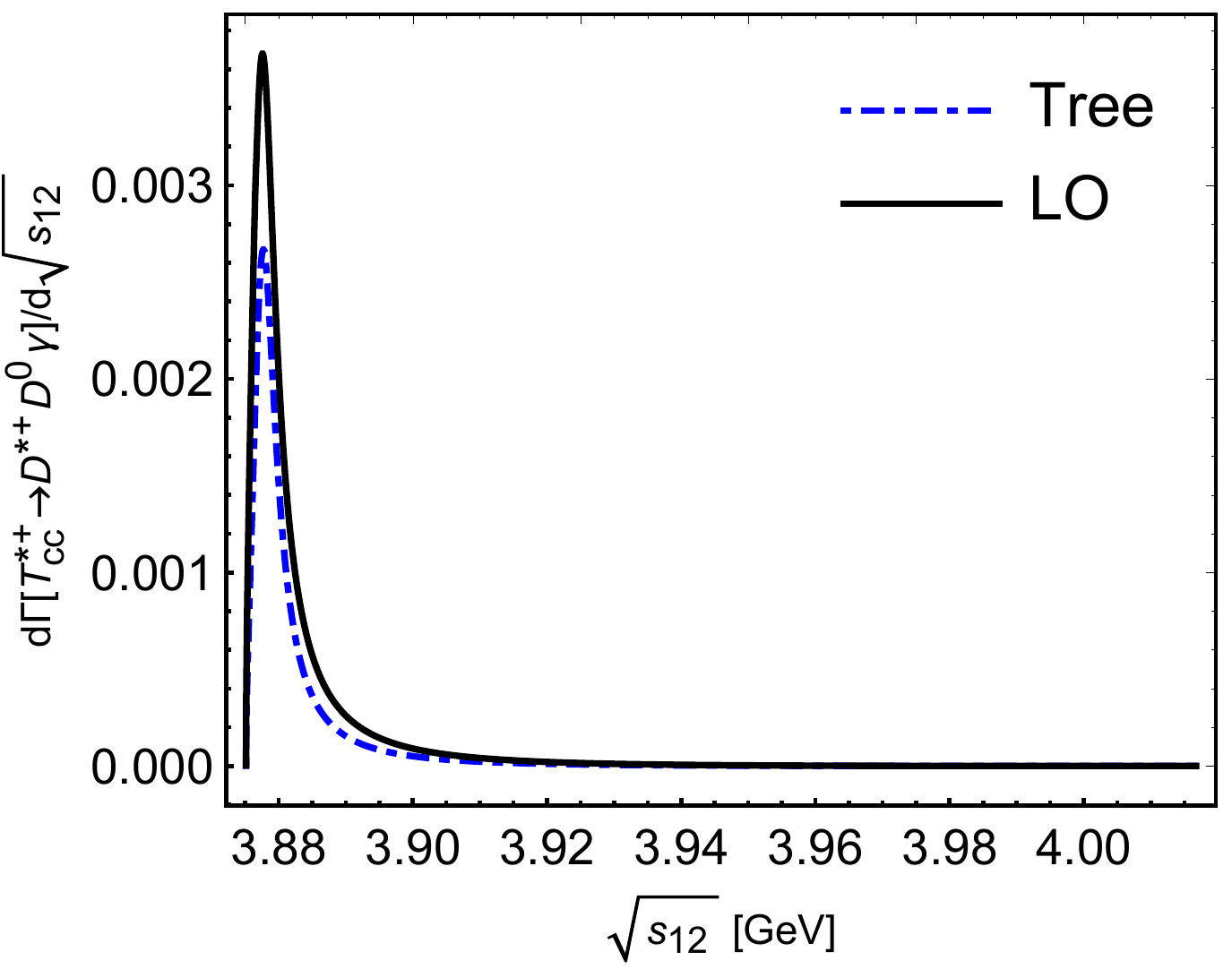}
    \label{fig_TsDscDg+}
    }
    \subfigure[$ T_{cc}^{\ast +}\to D^{\ast 0}D^+\gamma$] {\includegraphics[scale=0.355]{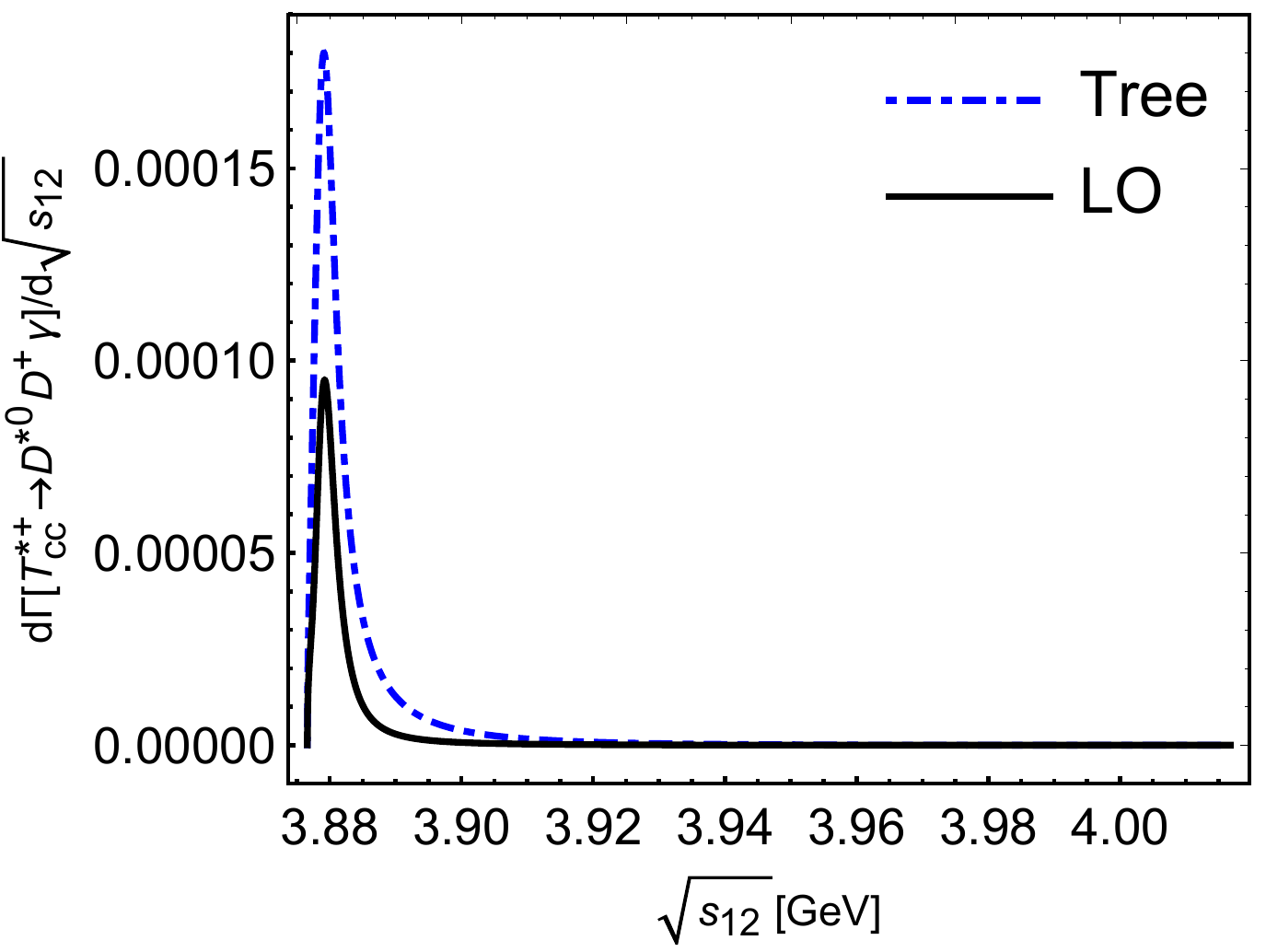} \label{fig_TsDsDcg-}
    }
   \caption{Differential decay rates for $T_{cc}^{\ast }\rightarrow D^{\ast} D \gamma$ with {$\sqrt{s_{12}}$ representing the $D^*D$ invariant mass.} The dashed and solid curves show the result including only the tree-level diagrams and that with the isoscalar $DD^*$ rescattering in addition, respectively.}
\label{fig_TsDsDgvss12}
\end{figure}

In this section, we present the partial decay widths for the decays $T_{cc}^{\ast} \to D^{\ast}D\gamma$, $T_{cc}\gamma$, and $T_{cc}\pi$.
In Table~\ref{Tab:3B2BDW}, we list the decay widths with the binding energy of the $T_{cc}^{*}$ being $\mathcal{B}=(503 \pm 40) \, \rm{keV}$~\cite{Du:2021zzh}. The second column denoted by $\Gamma_{\text{Tree}}$ lists the decay widths including only the contributions from the tree-level diagrams, and the third column marked by $\Gamma_{\rm{LO}}$ lists the LO decay widths including the tree-level and the $D^{\ast}D$ rescattering contributions. One sees that the isoscalar $D^{\ast}D$ rescattering which contains the $T_{cc}$ pole indeed contributes sizably, increases the tree-level results by about $50\%$ for $T_{cc}^{*+} \to D^{*+}D^0\gamma$, and decreases the tree-level results by about $58\%$ for $T_{cc}^{*+} \to D^{*0} D^+ \gamma$. To see the contributions of the $D^*D$ rescattering to the decay widths more clearly, the differential decay rates as a function of the $D^*D$ invariant mass $\sqrt{s_{12}}$ for $T_{cc}^{*+} \to D^{*+}D^{0}\gamma$ and $T_{cc}^{*+} \to D^{*0}D^{+}\gamma$ are shown in Fig.~\ref{fig_TsDsDgvss12}. One can clearly see the constructive interference and destructive interference effects for these two decays.
\begin{figure}[htbp]
    \subfigure[$T_{cc}^{\ast +}\to D^{\ast +}D^0\gamma$] {\includegraphics[scale=0.345]{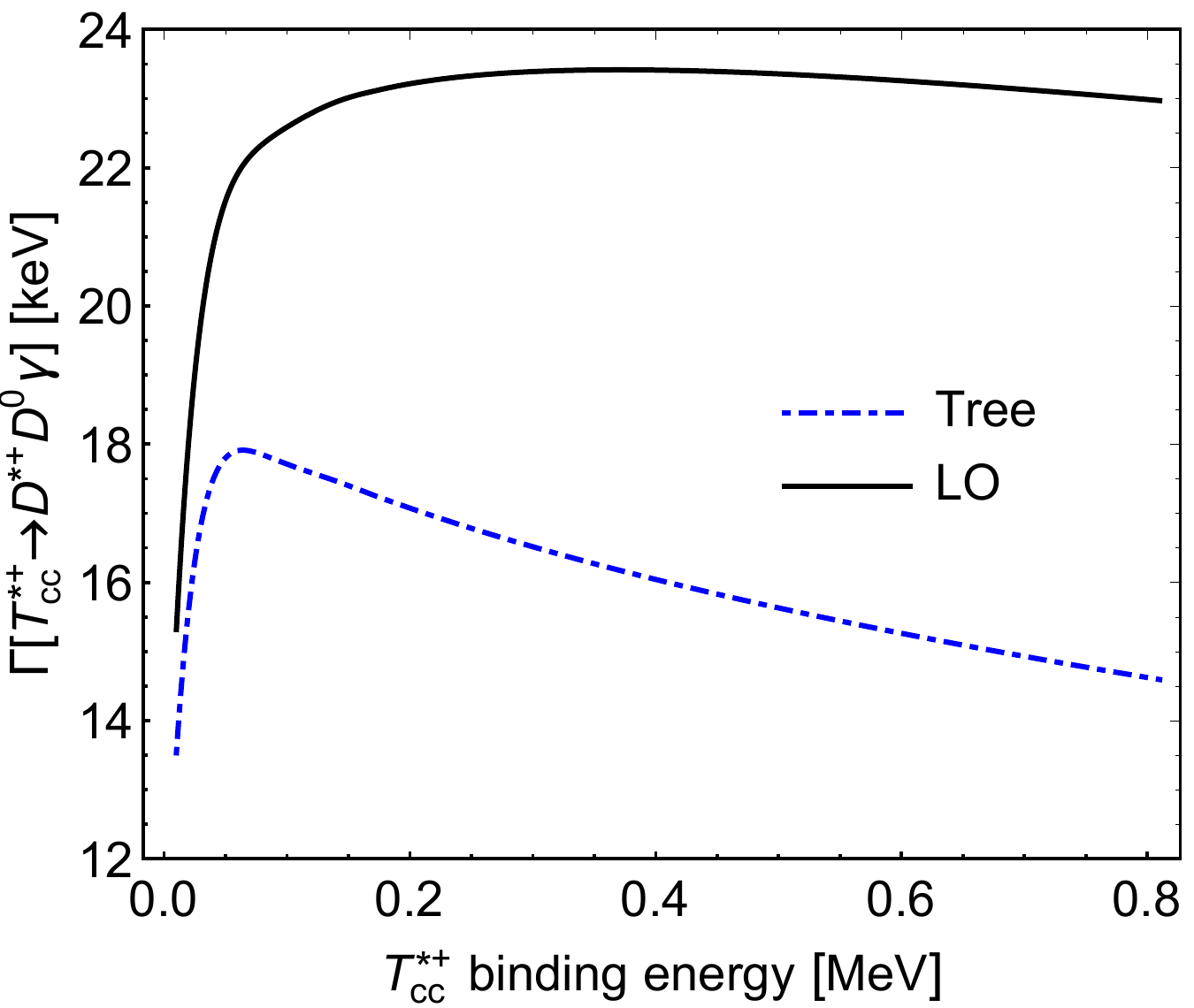}
    \label{fig_TsDscDgDWTreeLO}
    }
    \subfigure[$ T_{cc}^{\ast +}\to D^{\ast 0}D^+\gamma$] {\includegraphics[scale=0.355]{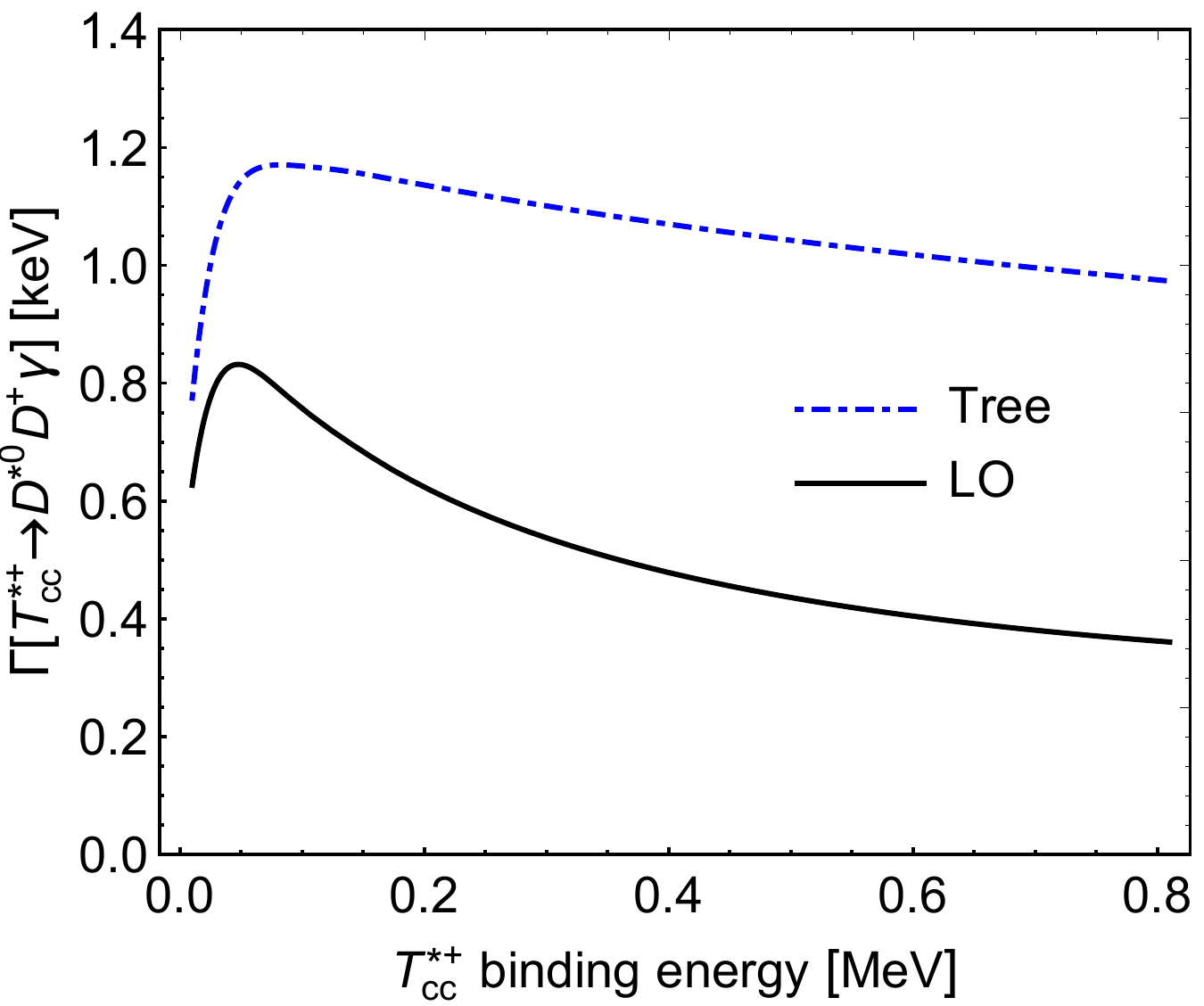} \label{fig_TsDsDcgDWTreeLO}
    }
   \caption{Partial decay widths of the $T_{cc}^{\ast }\rightarrow D^{\ast} D \gamma$ versus the binding energy of the $T_{cc}^{\ast +}$. Notations are the same as those in Fig.~\ref{fig_TsDsDgvss12}.}
\label{fig_TsDsDgDW}
\end{figure}

Since the binding energy of the $T_{cc}^{*}$ is uncertain, we further give the partial width of $T_{cc}^{\ast} \to D^{\ast}D\gamma$ with the binding energy varying from $0.01$ to $0.80~\rm{MeV}$ in Fig.~\ref{fig_TsDsDgDW},
where the blue dot-dashed lines show the decay widths from the tree-level diagram and the black solid-dashed lines show the LO decay width including both the tree-level and the $D^{\ast}D$ rescattering contributions. 

\begin{figure}[htbp]
    \subfigure[$T_{cc}^{\ast +}\to T_{cc}^+\gamma$] {\includegraphics[scale=0.45]{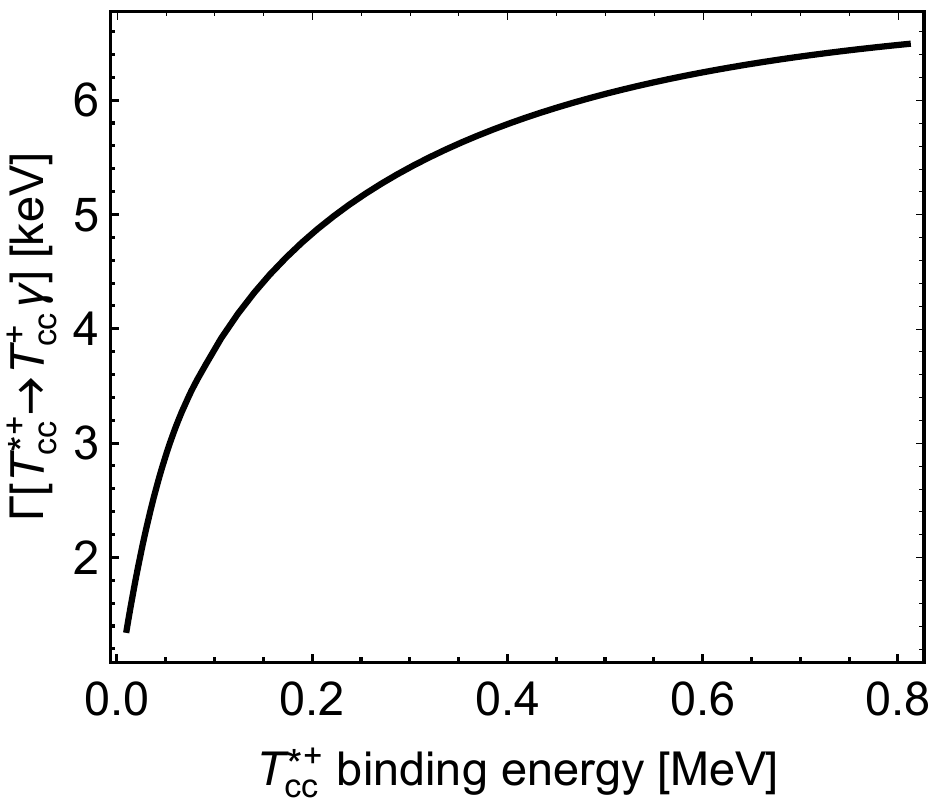} \label{fig_TsTcgDW}
    }
    \subfigure[$ T_{cc}^{\ast +}\to T_{cc}^+\pi^0$] {\includegraphics[scale=0.475]{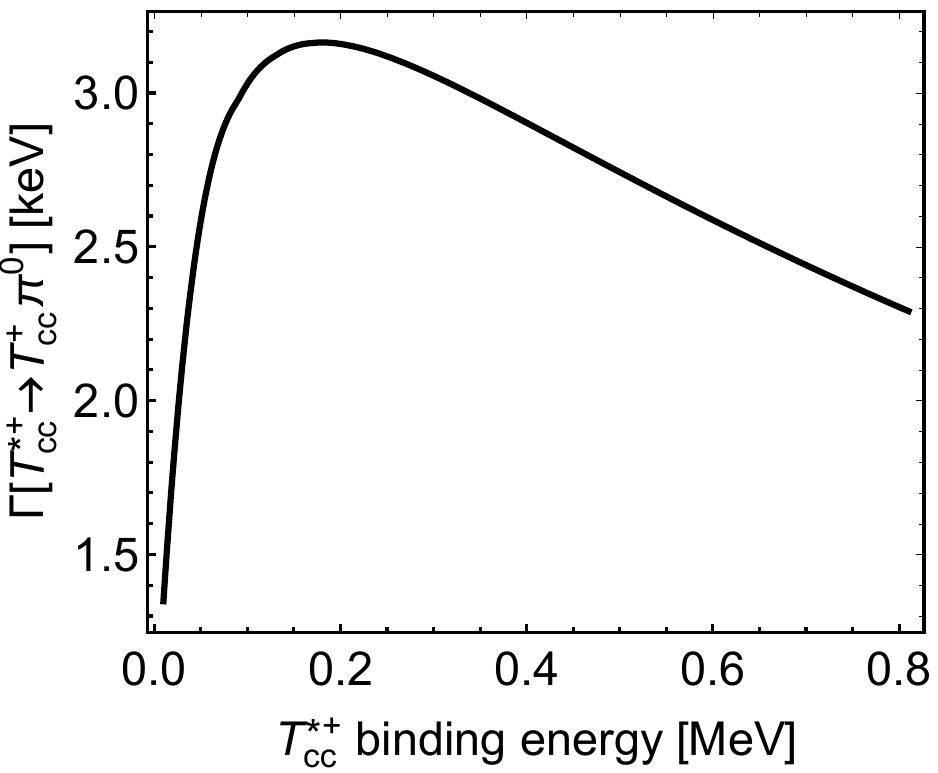} \label{fig_TsTcpiDW}
    }
   \caption{{Partial decay widths of the $T_{cc}^{\ast }\to T_{cc}\gamma$ and $T_{cc}\pi$ versus the binding energy of the $T_{cc}^{\ast}$.}}
    \label{Fig.TsTgpiDW}
\end{figure}

For the decays $T_{cc}^* \to T_{cc}\gamma$ and $T_{cc}\pi$, the decay widths with the binding energy of the $T_{cc}^*$ being $\mathcal{B}=(503 \pm 40)\, \rm{keV}$ are shown in the second column in Table~\ref{Tab:3B2BDW}, and the partial widths with the binding energy varying from $0.01$ to $0.80~\rm{MeV}$ are shown in Fig.~\ref{Fig.TsTgpiDW}. Here, we do not consider the unknown correlations between the binding energies of the $T_{cc}$ and $T_{cc}^*$.

\begin{figure}[htbp]
    \subfigure[] {\includegraphics[scale=0.35]{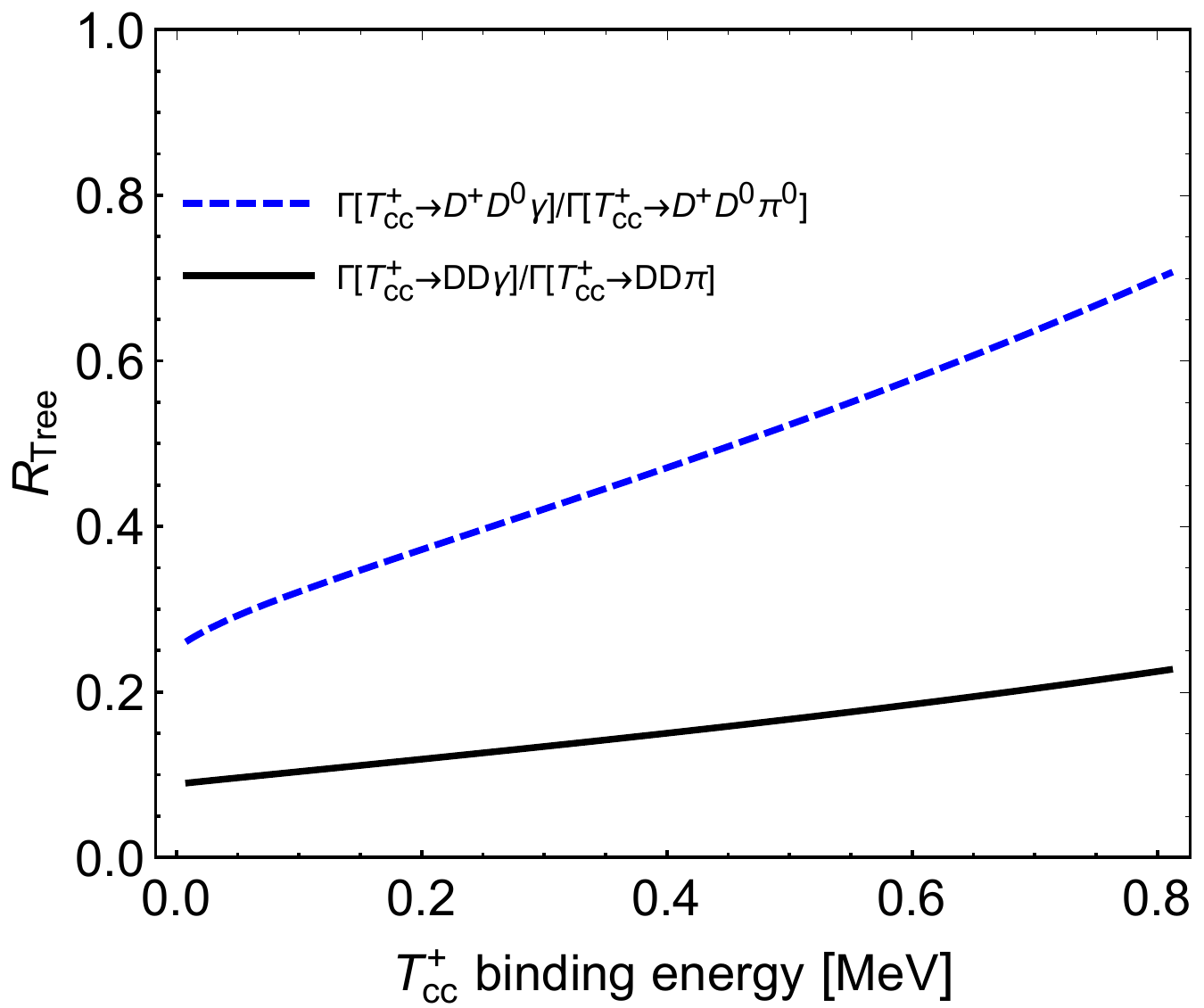} \label{fig_TR3BTree}
    }
    \subfigure[] {\includegraphics[scale=0.35]{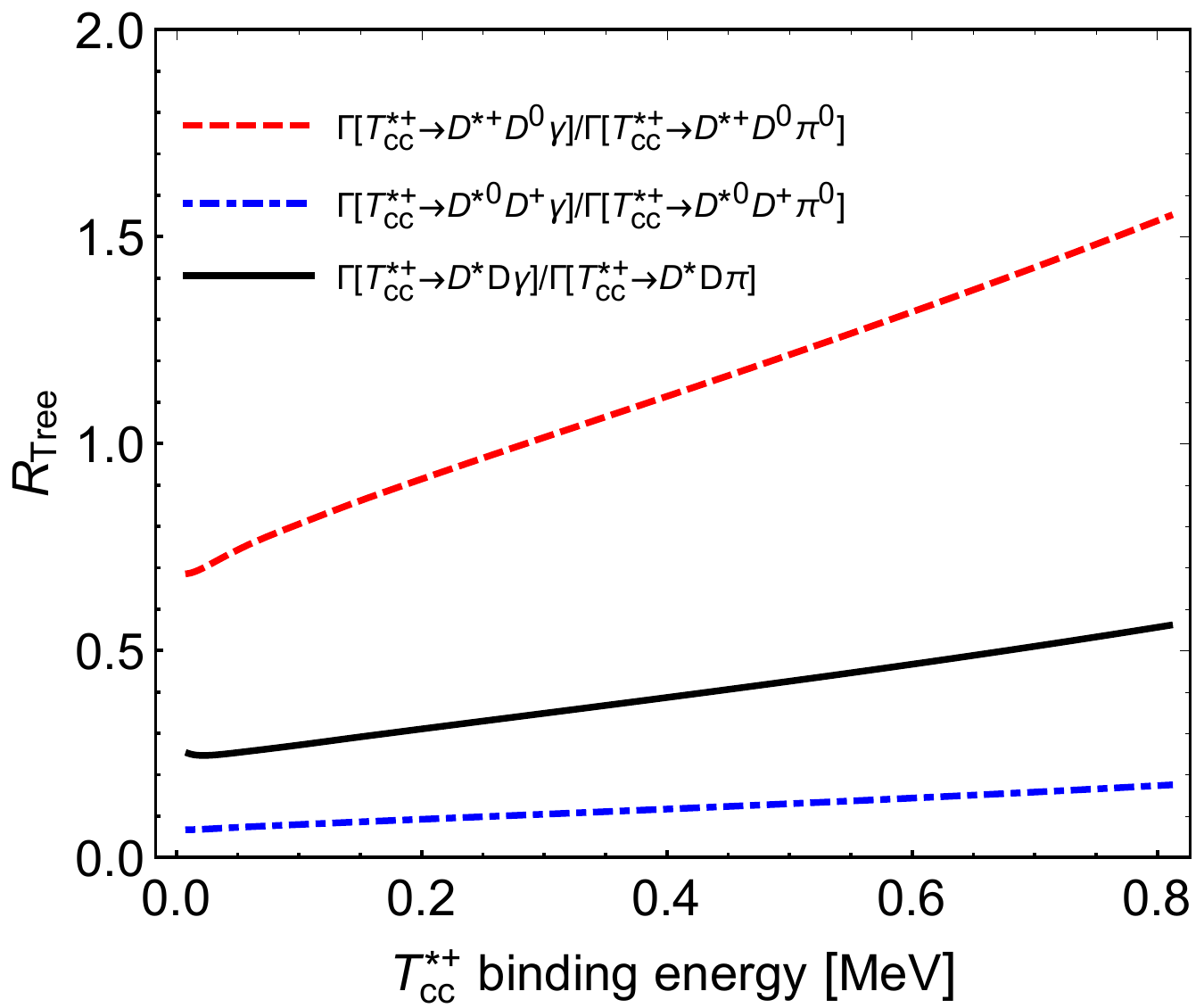} \label{fig_TsR3BTree}
    }\\
    \subfigure[] {\includegraphics[scale=0.35]{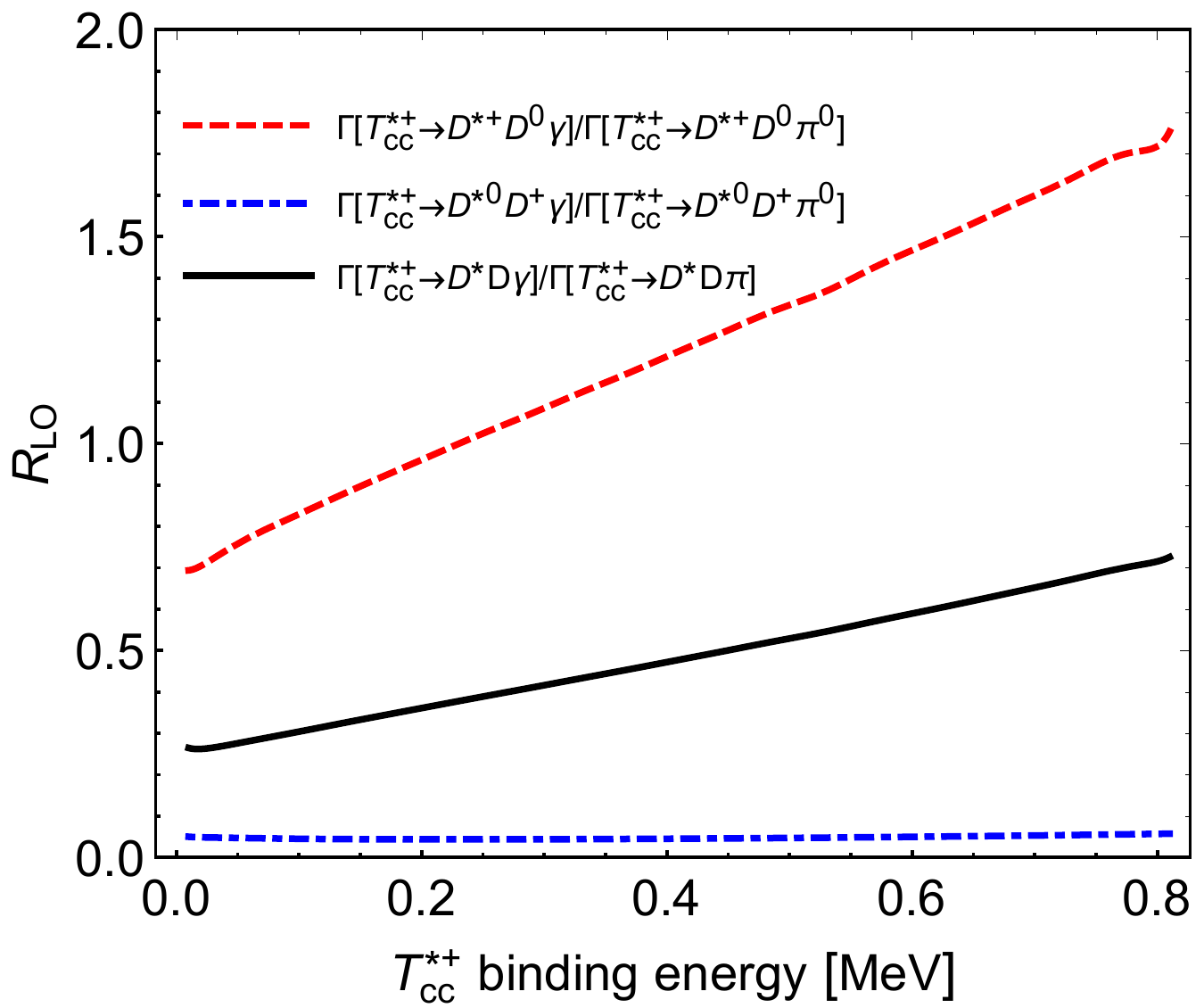} 
    \label{fig_TsR3BLO}
    }
    \subfigure[] {\includegraphics[scale=0.35]{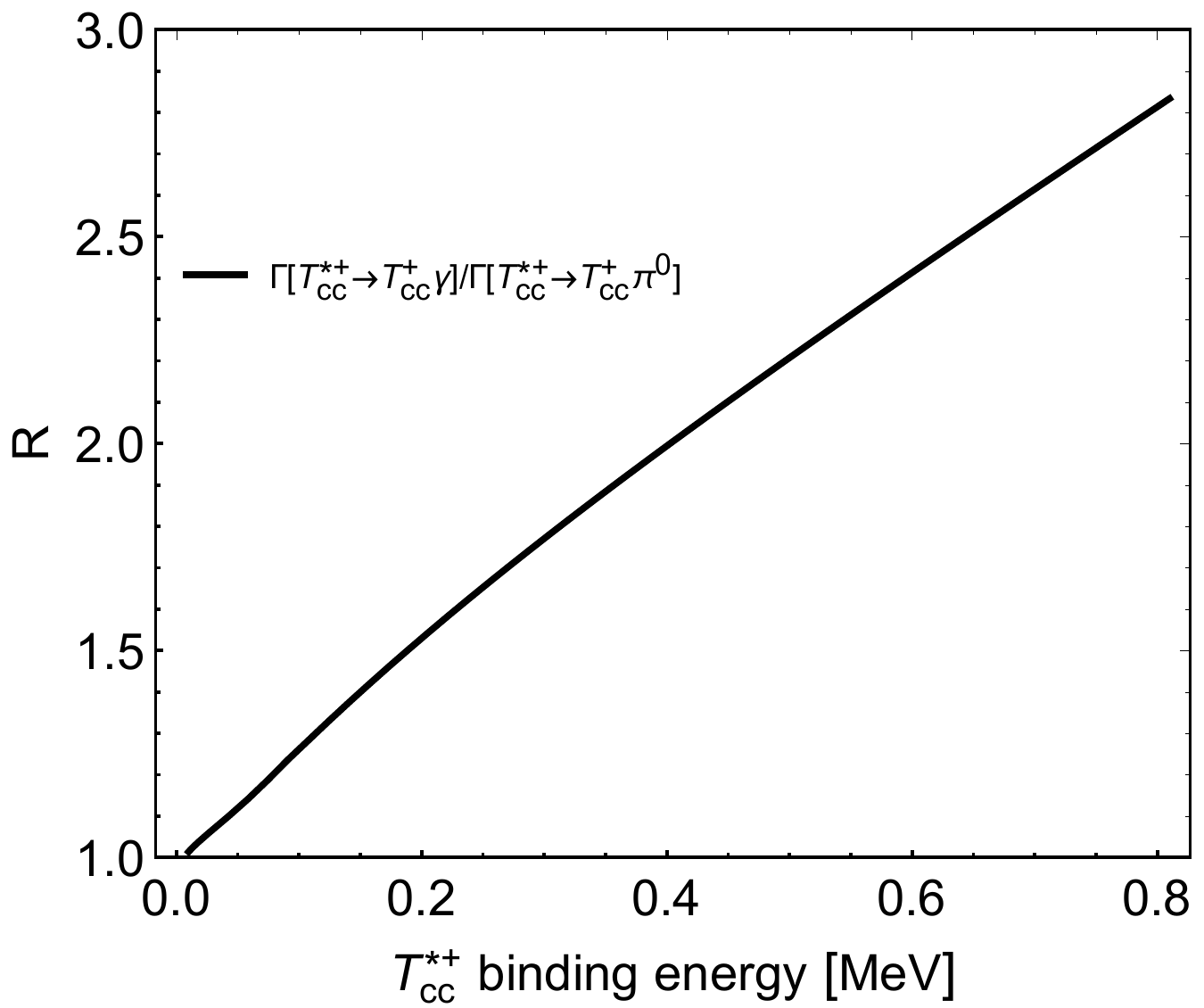} 
    \label{fig_TsR2B}
    }
   \caption{{Ratios $\Gamma[T_{cc}^{(*)} \to D^{(*)}D \gamma]/\Gamma[T_{cc}^{(*)} \to D^{(*)}D \pi]$ and $\Gamma(T_{cc}^* \to T_{cc}\gamma)/(T_{cc}^* \to T_{cc} \pi)$ versus the binding energy of $T_{cc}^{(*)}$. The results of $T_{cc}^+$ are obtained through revising Ref.~\cite{Fleming:2021wmk}. Here, the final states without electric charge indices mean the corresponding width is a sum over the modes related to each other with isospin symmetry; {\it e.g.}, $DD\pi$ means $D^+D^0\pi^0$ and $D^0D^0\pi^+$.}}
    \label{Fig.BR}
\end{figure}

Combining the hadronic and radiative decay widths of the $T_{cc}^+$ calculated in Ref.~\cite{Fleming:2021wmk} and the hadronic decay widths of the $T_{cc}^{*+}$ in Ref.~\cite{Jia:2022qwr}, the ratios $\Gamma[T_{cc}^{(*)} \to D^{(*)}D\gamma]/\Gamma[T_{cc}^{(*)} \to D^{(*)}D\pi]$ and $\Gamma[T_{cc}^* \to T_{cc}\gamma]/\Gamma[T_{cc}^* \to T_{cc}\pi]$ are depicted in Fig.~\ref{Fig.BR} to show the relations among different channels. 

Summing up these three-body partial decay widths of $T_{cc}^*\to D^*D\gamma$ leads
the total radiative decay width of the $T_{cc}^{*+}$ to be
\begin{equation}
    \Gamma_{\gamma}(T_{cc}^{*+}) = (23.8 \pm 0.1)~{\rm keV}. \label{eq:total}
\end{equation}
In Ref.~\cite{Jia:2022qwr}, the obtained hadronic decay width of the $T_{cc}^*$ is about $(41\pm 2)$ keV, so the predicted 
total width of the $T_{cc}^*$ is 
\begin{equation}
    \Gamma(T_{cc}^{*+}) = (65 \pm 2)~{\rm keV},
\end{equation}
which is larger than that of the $T_{cc}$ and can be regarded as the main result of our work.

\section{Partial decay widths for $T_{cc}^{*} \to DD\gamma\gamma$ and $DD\gamma\pi$}\label{sec:Results4B}

In the three-body decay $T_{cc}^{*}\to D^*D\gamma$, the resonant $D^*$ in the final states can further decay into the $D\gamma$ or $D\pi$, so the $T_{cc}^*$ will decay into the stable final states $D^+ D^0\gamma \gamma$, $D^+ D^0\gamma\pi^0$, and $D^0 D^0 \gamma \pi^+$. Since the $D^{*+} D^0 \gamma$ intermediate state can decay into the same $D^+D^0\gamma\gamma$ final states as $D^{*0} D^+ \gamma$, the same $D^+D^0\gamma\pi^0$ final states as $D^{*+} D^0 \pi^0$ and $D^{*0} D^+ \pi^0$, and the same $D^0D^0\gamma\pi^+$ final states as $D^{*0}D^0\pi^+$, it interferes with these processes. In the following, we will calculate the decay widths of the $T_{cc}^{*+} \to D^+ D^0\gamma \gamma$, $D^+ D^0\gamma\pi^0$, and $D^0 D^0\gamma\pi^+$ to show that the interference between the intermediate three-body states is small, and it is a good approximation that we consider only the three-body $D^{*}D\gamma$ final states to calculate the $T_{cc}^{*}$ radiative decay width.

\begin{figure}[htbp]
    \subfigure[] {
        \includegraphics[scale=1.2]{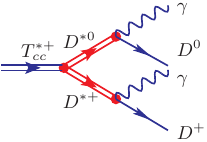}
        \label{fig_TsDcDggFDa}
    }
    \subfigure[] {
        \includegraphics[scale=1.2]{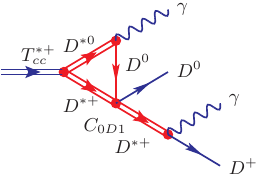}
        \label{fig_TsDcDggFDb}
    }
    \subfigure[] {
        \includegraphics[scale=1.2]{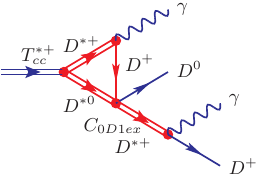}
        \label{fig_TsDcDggFDc}
    }
    \subfigure[] {
        \includegraphics[scale=1.2]{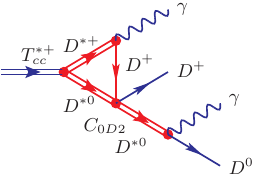}
        \label{fig_TsDcDggFDd}
    }
    \subfigure[] {
        \includegraphics[scale=1.2]{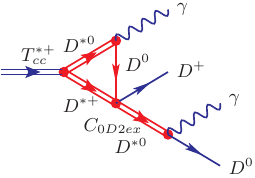}
        \label{fig_TsDcDggFDe}
    }
    \caption{Feynman diagrams for calculating the partial decay width of $T_{cc}^{*+}\rightarrow D^{+}D^0\gamma\gamma$. The double lines represent the spin-1 mesons, $T_{cc}^{*+}$, $D^{*0}$ and $D^{*+}$; the single thin lines represent the pseudoscalar charmed mesons, $D^+$ and $D^0$; and the wavy lines represent the photon.}
    \label{fig_TsDcDggFD}
\end{figure}

\begin{figure}[htbp]
    \subfigure[] {
        \includegraphics[scale=1.1]{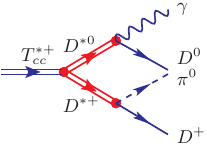}
        \label{fig_TsDcDpi0gFDa}
    }
    \subfigure[] {
        \includegraphics[scale=1.1]{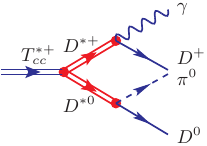}
        \label{fig_TsDcDpi0gFDb}
    }
    \subfigure[] {
        \includegraphics[scale=1.1]{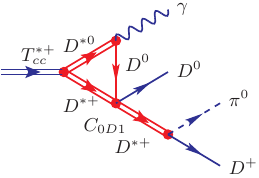}
        \label{fig_TsDcDpi0gFDc}
    }
    \subfigure[] {
        \includegraphics[scale=1.1]{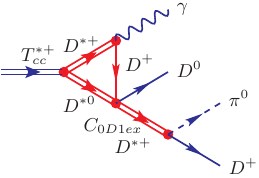}
        \label{fig_TsDcDpi0gFDd}
    }
    \subfigure[] {
        \includegraphics[scale=1.1]{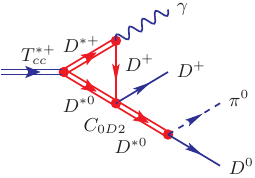}
        \label{fig_TsDcDpi0gFDe}
    }
     \subfigure[] {
        \includegraphics[scale=1.1]{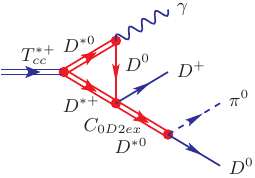}
        \label{fig_TsDcDpi0gFDf}
    }\\
     \subfigure[] {
        \includegraphics[scale=1.1]{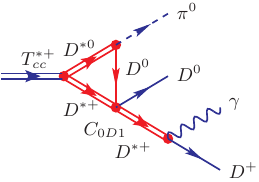}
        \label{fig_TsDcDpi0gFDg}
    }
     \subfigure[] {
        \includegraphics[scale=1.1]{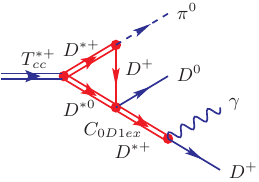}
        \label{fig_TsDcDpi0gFDh}
    }
     \subfigure[] {
        \includegraphics[scale=1.1]{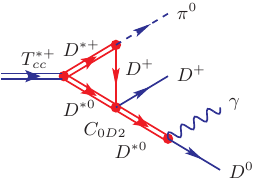}
        \label{fig_TsDcDpi0gFDi}
    }\\
     \subfigure[] {
        \includegraphics[scale=1.1]{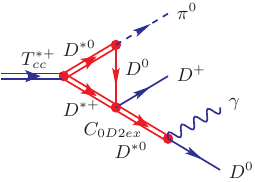}
        \label{fig_TsDcDpi0gFDj}
    }
     \subfigure[] {
        \includegraphics[scale=1.1]{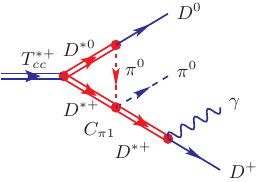}
        \label{fig_TsDcDpi0gFDk}
    }
     \subfigure[] {
        \includegraphics[scale=1.1]{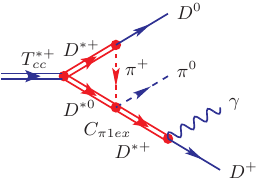}
        \label{fig_TsDcDpi0gFDl}
    }\\
     \subfigure[] {
        \includegraphics[scale=1.1]{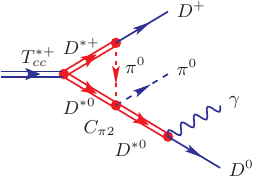}
        \label{fig_TsDcDpi0gFDm}
    }
     \subfigure[] {
        \includegraphics[scale=1.1]{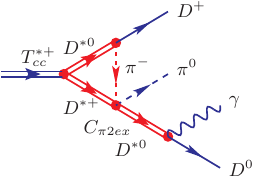}
        \label{fig_TsDcDpi0gFDn}
    }
    \caption{Feynman diagrams for calculating the partial decay width of $T_{cc}^{*+}\rightarrow D^{+}D^0\gamma\pi^0$. The double lines represent the spin-1 mesons, $T_{cc}^{*+}$, $D^{*0}$, and $D^{*+}$; the single thin lines represent the pseudoscalar charmed mesons, $D^+$ and $D^0$; the wavy lines represent the photon; and the dashed lines represent the pion.}
    \label{fig_TsDcDpigFD}
\end{figure}

\begin{figure}[htbp]
    \subfigure[] {
        \includegraphics[scale=1.2]{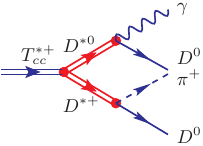}
        \label{fig_TsDDpicgFDa}
    }
    \subfigure[] {
        \includegraphics[scale=1.2]{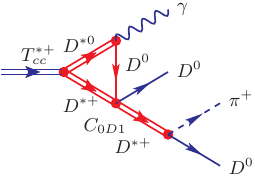}
        \label{fig_TsDDpicgFDb}
    }
    \subfigure[] {
        \includegraphics[scale=1.2]{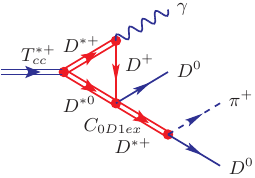}
        \label{fig_TsDDpicgFDc}
    }
    \subfigure[] {
        \includegraphics[scale=1.2]{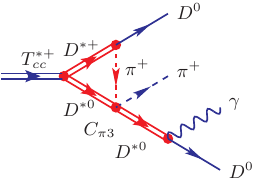}
        \label{fig_TsDDpicgFDd}
    }
    \subfigure[] {
        \includegraphics[scale=1.2]{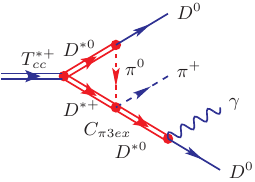}
        \label{fig_TsDDpicgFDe}
    }
    \caption{Feynman diagrams for calculating the partial decay width of $T_{cc}^{*+}\rightarrow D^{+}D^0\gamma\pi^+$. The double lines represent the spin-1 mesons, $T_{cc}^{*+}$, $D^{*0}$, and $D^{*+}$; the single thin lines represent the pseudoscalar charmed mesons, $D^+$ and $D^0$; the wavy lines represent the photon; and the dashed lines represent the pion.}
    \label{fig_TsDDpicgFD}
\end{figure}

The diagrams for the four-body decays $T_{cc}^{*+} \to D^+ D^0\gamma \gamma $, $T_{cc}^{*+} \to D^+ D^0 \gamma \pi^0$, and $T_{cc}^{*+}\to D^0 D^0 \gamma \pi^+$ are shown in Figs.~\ref{fig_TsDcDggFD}, \ref{fig_TsDcDpigFD}, and \ref{fig_TsDDpicgFD}, respectively. The amplitudes for all the diagrams are collected in Appendix~\ref{Appendix:4body_amplitudes}.  The  four-body decay rate is given by
\begin{align}
d\Gamma_{\text{4B}}=2M \prod_{i=1}^{n}  2E_i \frac{1}{2SM}\frac{1}{2j+1} \sum_{\text{spins}} \left\vert \mathcal{A}_{\text{4B}}\right\vert^2 d\Phi_4,
\label{Eq.4-body_decay_rate}
\end{align}
where the overall factor comes from the normalization of nonrelativistic particles and $E_i$ ($i=1,\ldots, n$, $n\leq 4$ is the number of final-state nonrelativistic particles) are the energies of the final-state particles in the $T_{cc}^{*}$ rest frame. $S$ is the symmetry factor accounting for identical particles in the final state---$S=2$ for the $T_{cc}^{*+} \to D^+ D^0\gamma \gamma/D^0 D^0 \gamma \pi^+ $ decays and $S=1$ for the $T_{cc}^{*+} \to D^+ D^0 \gamma \pi^0$ decay.
The four-body phase space in Eq.~\eqref{Eq.4-body_decay_rate} reads (for details of the derivations, see Refs.~\cite{Jing:2020tth,Jia:2022qwr})
\begin{align}
d\Phi_4(P;p_1,...,p_4)= \frac{1}{(8\pi^2)^4M} \int_{m_1+m_2}^{M-m_3-m_4}d\sqrt{s_{12}}\int_{m_3+m_4}^{M-\sqrt{s_{12}}}d\sqrt{s_{34}} \int d\Omega_1^*d\Omega_3^{\prime}d\Omega \vert \vec{p}_1^{\,*} \vert \vert \vec{p}_3^{\, \prime} \vert \vert \vec{q} \, \vert,
\label{Eq.Phi4}
\end{align}
where $s_{12}=(p_1+p_2)^2$, $s_{34}=(p_3+p_4)^2$, $\vec{q} = (|\vec{q}|, \Omega) $ is the three-momentum of the $(1,2)$ particle system in the rest frame of the initial particle $T_{cc}^*$, $\vec{p}^{\,*}_{1}=(|\vec{p}^{\,*}_{1}|, \Omega_{1}^*)$ is the three-momentum of particle 1 in the c.m. frame of the $(1,2)$ particle system, and $\vec{p}^{\,\prime}_3=(|\vec{p}^{\,\prime}_3|, \Omega_3')$ is the three-momentum of particle 3 in the c.m. frame of the $(3,4)$ particle system. 
The magnitudes of the three-momenta are given by
\begin{align}
  \vert \vec{q} \, \vert= \frac{\lambda^{1/2}(M^2,s_{12},s_{34})}{2M}, \quad
  \vert \vec{p}_1^{\,*} \vert= \frac{\lambda^{1/2}(s_{12}, m_1^2, m_2^2)}{2\sqrt{s_{12}}}, \quad
  \vert \vec{p}_3^{\,\prime} \vert=&\, \frac{\lambda^{1/2}(s_{34}, m_3^2, m_4^2)}{2\sqrt{s_{34}}},
\end{align}
with $\lambda(x,y,z)\equiv x^2+y^2+z^2-2(xy+xz+yz)$ being the K\"all\'en triangle function. 

The differential decay rate for the $T_{cc}^{*+} \to DD\gamma\gamma$ at LO including the $D^*D$ FSI reads
\begin{align}
\frac{d\Gamma[T_{cc}^{*} \to DD\gamma\gamma]}{d\sqrt{s_{12}}d\sqrt{s_{34}}}=&\, 2m_{T_{cc}^*} 2p_2^0 2p_4^0 \frac{1}{4m_{T_{cc}^*}} \frac{1}{3}  \frac{1}{(8\pi^2)^4 m_{T_{cc}^*}} d\Omega_1^{\, *}d\Omega_3^{\, \prime}d\Omega \vert \vec{p}_{1}^{\, *} \vert \vert \vec{p}_{3}^{{\, \prime}} \vert \vert \vec{q} \, \vert \sum_{\text{spins}} \vert \mathcal{A}_{\text{LO}} \vert^2,
\end{align}
where “$1$,” “$2$,” “$3$,” and “$4$” denote the $\gamma$, $D^0$, $\gamma$, and $D^+$ particles, respectively, $p_2^0$ and $p_4^0$ are the energies of the $D^0$ and $D^+$ mesons in the $T_{cc}^*$ rest frame, respectively, $\vec{p}_1^{\, *}$ and $\vec{p}_3^{\, \prime}$ are the three-momenta of the two final-state photons in the c.m. frame of the $(1, 2)$ and $(3, 4)$ two-particle systems, respectively, and $\mathcal{A}_{\text{LO}}$ is the LO amplitude including the contribution from the tree-level and $D^*D$ rescattering diagrams.

The differential decay rate for the $T_{cc}^{*+} \to DD\gamma\pi$ up to NLO including the $D^*D$ and $D^*\pi$ rescattering corrections reads
\begin{align}
\frac{d\Gamma[T_{cc}^{*} \to DD\gamma\pi]}{d\sqrt{s_{12}}d\sqrt{s_{34}}}=&\, 2m_{T_{cc}^*} 2\tilde{p}_2^0 2\tilde{p}_3^0 2\tilde{p}_4^0 \frac{1}{4m_{T_{cc}^*}} \frac{1}{3} \frac{1}{(8\pi^2)^4 m_{T_{cc}^*}} d\Omega_1^{\, *}d\Omega_3^{\, \prime}d\Omega \vert \vec{p}_{1}^{\, *} \vert \vert \vec{p}_{3}^{{\, \prime}} \vert \vert \vec{q}\, \vert \nonumber\\
&\, \times \left\{\sum_{\text{spins}} \vert \mathcal{A}_{\text{LO}} \vert^2+2\text{Re}\left[\sum_{\text{spins}}\mathcal{A}_{\text{LO}} \times \mathcal{A}_{\text{NLO}}\right]\right\},
\end{align}
where “$1$,” “$2$,” “$3$,” and “$4$” denote the $\gamma$, $D^0$, $\pi^0$ ($\pi^+$) and $D^+$ ($D^0$) particles, respectively, for the $T_{cc}^{*+} \to D^+D^0\gamma\pi^0$ ($D^0D^0\gamma\pi^+$) decay, $\tilde{p}_3^0$ and $\tilde{p}_4^0$ are the energies of $\pi^0$ ($\pi^+$) and $D^+$($D^0$) in the $T_{cc}^*$ rest frame, respectively, and $\mathcal{A}_{\text{NLO}}$ is the NLO amplitude including only the $D^*\pi$ rescattering diagrams. The second term in the curly brackets includes the correction of the $D^*\pi$ rescattering, which is the interference term between the LO and NLO amplitudes.

\begin{table}[bth]
\caption{\label{Tab:Tccstar+4Body} Partial decay widths of the $T_{cc}^{\ast } \to DD\gamma\gamma$ and $DD\gamma\pi$ for $T_{cc}^*$ with a binding energy $\mathcal{B}=(503 \pm 40)\, \rm{keV}$. The second column lists results from the tree-level diagrams, the third column gives the LO decay widths including contributions from both the tree-level and $D^{\ast}D$ rescattering diagrams, and the last column lists the final results including corrections from the $D^{\ast}\pi$ rescattering to $\Gamma_{\text{LO}}$. There are no $D^*\pi$ scatterings in the decays of $T_{cc}^* \to DD\gamma\gamma$.}
\renewcommand{\arraystretch}{1.2}
\setlength{\tabcolsep}{20pt}{
\begin{tabular*}{\columnwidth}{l|c|c|c}
\hline\hline                                  
        $\Gamma$[keV]
        &$\text{Tree}$
        &$\text{LO}$ 
        &$\text{NLO}$
        \\[3pt]        
\hline
 $\Gamma[T_{cc}^{\ast+}\rightarrow D^+ D^0\gamma \gamma ] $ &$0.5 \pm 0.0$ &$0.5 \pm 0.0$ & / \\[3pt]
 \hline 
 $\Gamma[T_{cc}^{\ast+}\rightarrow D^{*+}D^0 \gamma] \times \text{Br}(D^{*+} \rightarrow D^+\gamma)$ &\multirow{2}{*}{$0.6 \pm 0.0$} &\multirow{2}{*}{$0.5 \pm 0.0$} &\multirow{2}{*}{/}\\[3pt]
 +$\Gamma[T_{cc}^{\ast+}\rightarrow D^{*0} D^+\gamma] \times \text{Br}(D^{*0} \rightarrow D^0 \gamma)$ & & \\ [3pt] 
\hline
 $\Gamma[T_{cc}^{\ast+}\rightarrow D^+ D^0 \gamma \pi^0]$ &$8.5 \pm 0.2$ &$10.8^{+0.2}_{-0.1}$ &$10.7^{+0.2}_{-0.1}$\\[3pt]
 \hline
 $\Gamma[T_{cc}^{\ast+}\rightarrow D^{*+} D^0\gamma] \times \text{Br}(D^{*+} \rightarrow D^+\pi^0)$ &\multirow{4}{*}{$8.5 \pm 0.2$} &\multirow{4}{*}{$11.0 \pm 0.2$} &\multirow{4}{*}{$10.6 \pm 0.2$}\\[3pt]
 +$\Gamma[T_{cc}^{\ast+}\rightarrow D^{*0}D^+\gamma] \times \text{Br}(D^{*0}\rightarrow D^0\pi^0)$ & & \\ [3pt] 
 +$\Gamma[T_{cc}^{\ast+}\rightarrow D^{*+}D^0\pi^0] \times \text{Br}(D^{*+}\rightarrow D^+\gamma)$ & &
 \\ [3pt] 
 +$\Gamma[T_{cc}^{\ast+}\rightarrow D^{*0}D^+\pi^0] \times \text{Br}(D^{*0}\rightarrow D^0\gamma)$ & &
 \\ [3pt]
 \hline
 $\Gamma[T_{cc}^{\ast+}\rightarrow D^0D^0\gamma\pi^+]$ &$15.2^{+0.5}_{-0.1}$ &$19.5^{+0.5}_{-0.2}$ &$19.3^{+0.5}_{-0.2}$\\[3pt]
 \hline
 $\Gamma[T_{cc}^{\ast+}\rightarrow D^{*+} D^0\gamma] \times \text{Br}(D^{*+} \rightarrow D^0\pi^+)$ &\multirow{2}{*}{$17.0 \pm 0.4$} &\multirow{2}{*}{$22.2 \pm 0.3$} &\multirow{2}{*}{$22.0 \pm 0.3$}\\[3pt]
 +$\Gamma[T_{cc}^{\ast+}\rightarrow D^{*0}D^0\pi^+] \times \text{Br}(D^{*0}\rightarrow D^0\gamma)$ & & \\ [3pt]
\hline\hline
\end{tabular*}}
\end{table}
\begin{figure}[htbp]
    \subfigure[$T_{cc}^{\ast +}\to D^+ D^0\gamma \gamma $] {\includegraphics[scale=0.23]{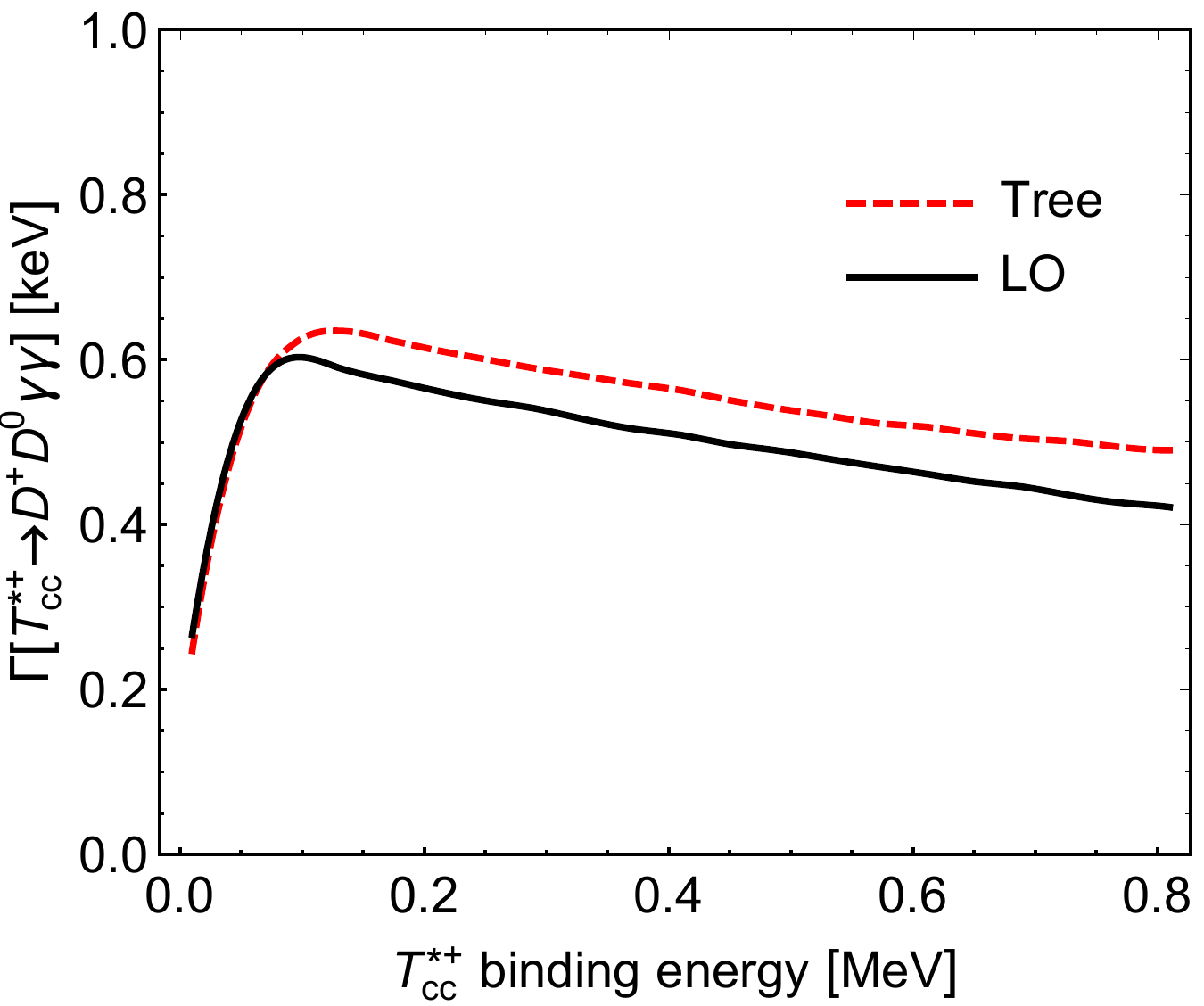} \label{fig_TsDcDggTreeLO}
    }
    \subfigure[$T_{cc}^{\ast +}\to D^+ D^0 \gamma\pi^0$] {\includegraphics[scale=0.23]{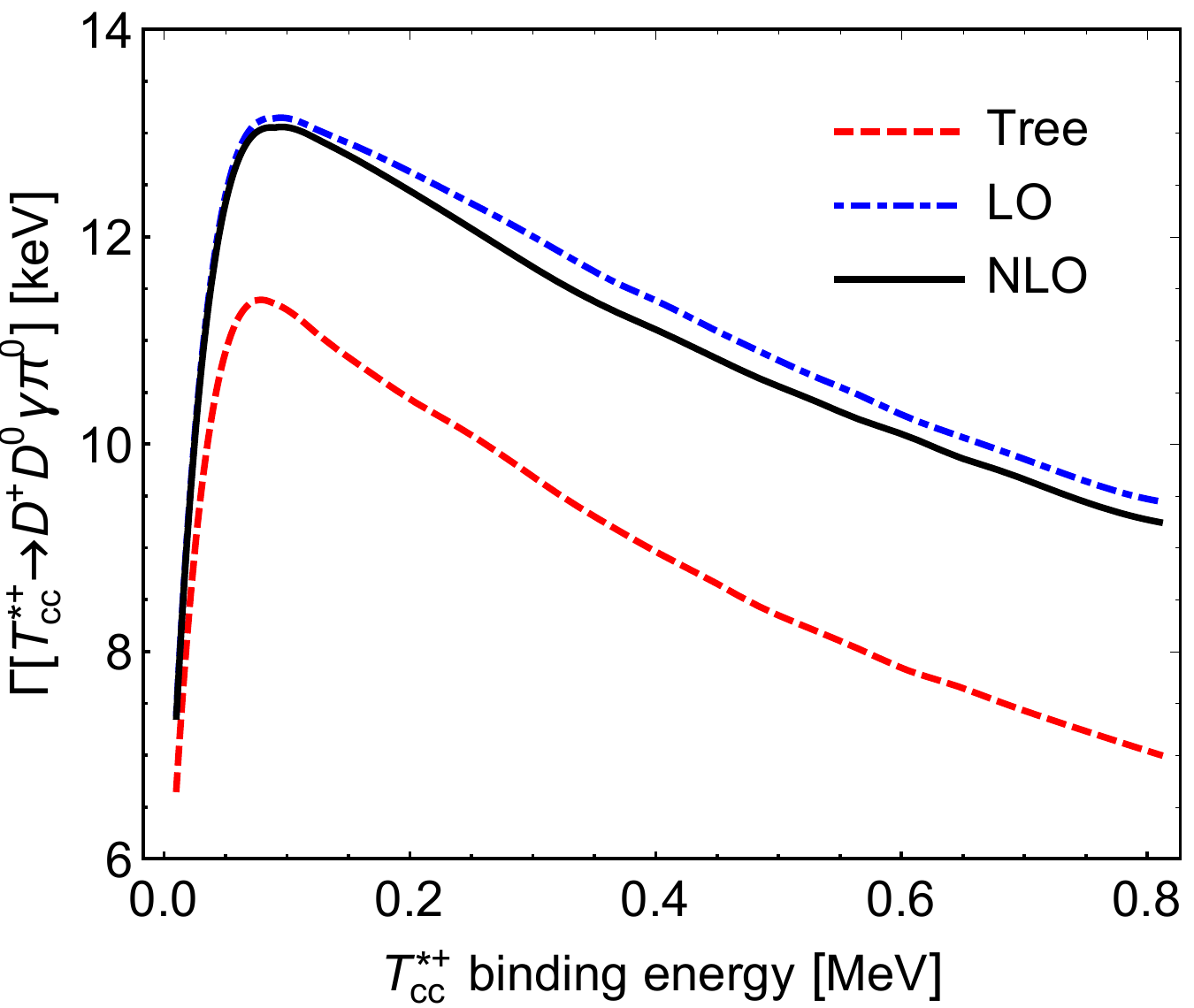} \label{fig_TsDcDpi0gTreeLONLO}
    }
    \subfigure[$T_{cc}^{\ast +}\to D^0 D^0\gamma\pi^+ $] {\includegraphics[scale=0.23]{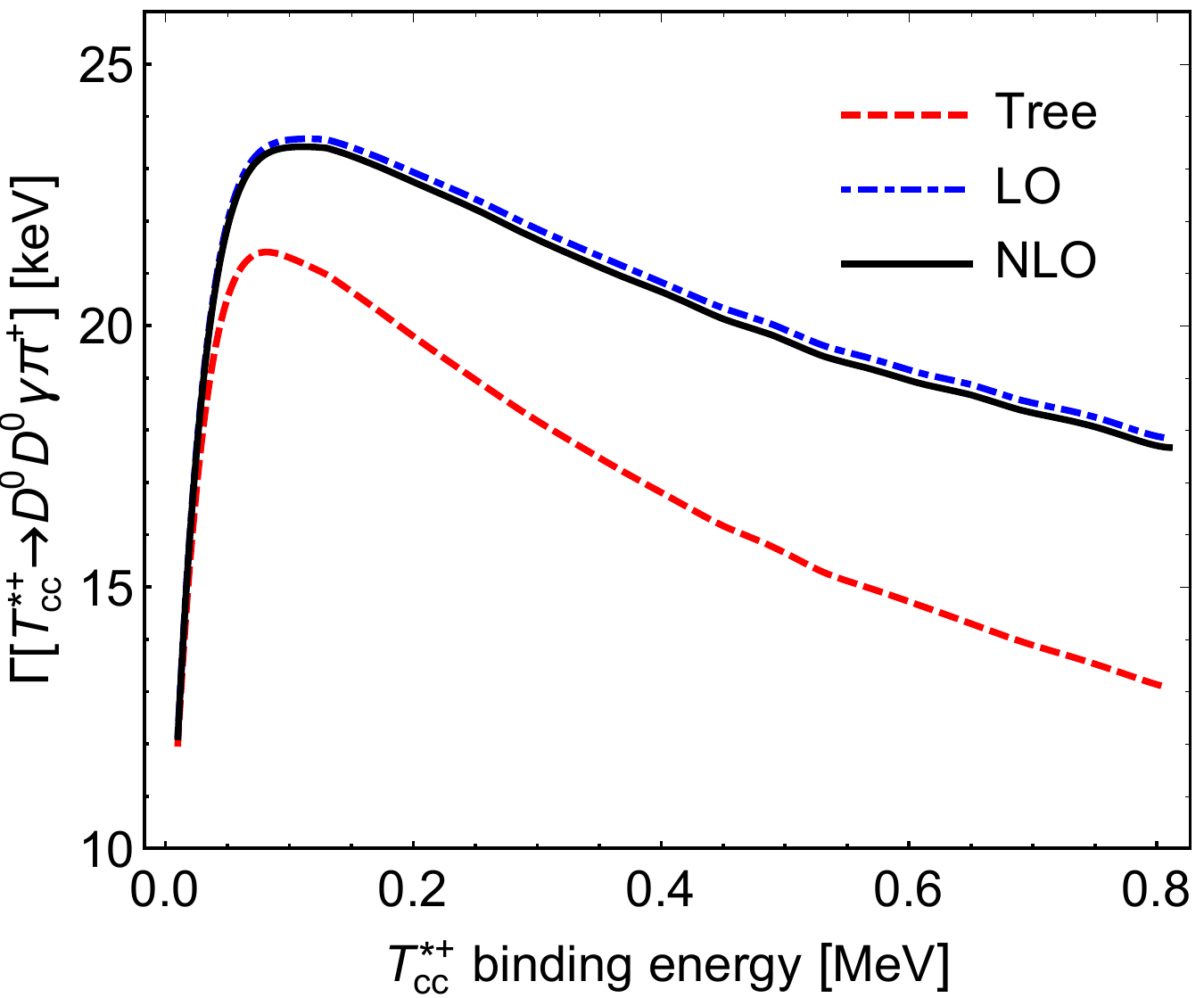} \label{fig_TsDcDpi0gTreeLONLO}
    }
   \caption{Partial decay widths of the $T_{cc}^{\ast }\rightarrow DD\gamma\gamma$ and $DD\gamma\pi$ versus the binding energy of the $T_{cc}^{\ast +}$.}
    \label{fig_TsDDg/pigDW}
\end{figure}
\begin{figure}[htbp]
    \subfigure[$T_{cc}^{\ast +}\to D^+ D^0\gamma \gamma $] {\includegraphics[scale=0.23]{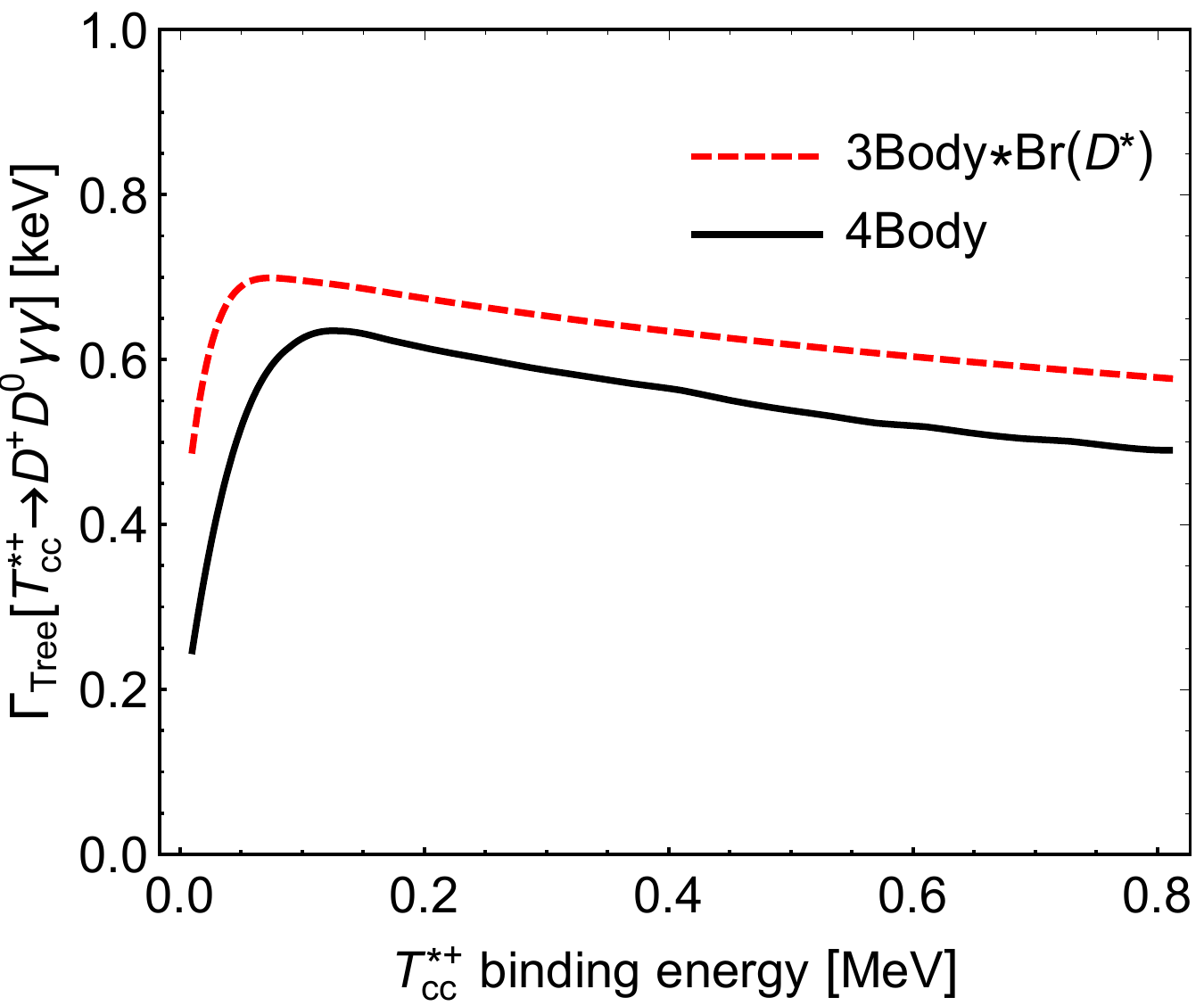} \label{fig_TsDcDgg3B4BTree}
    }
    \subfigure[$T_{cc}^{\ast +}\to D^+ D^0\gamma \gamma $] {\includegraphics[scale=0.23]{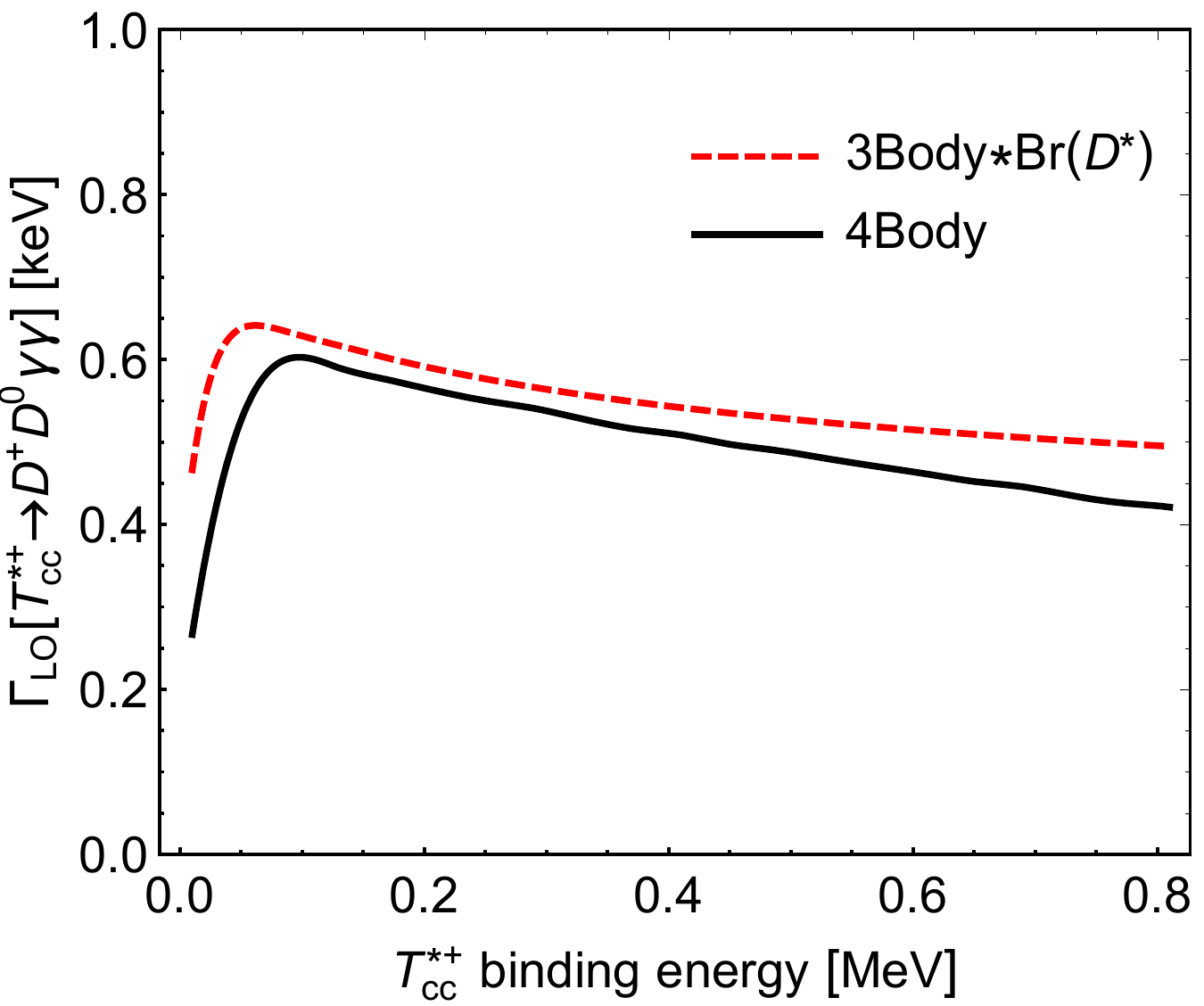} \label{fig_TsDcDgg3B4BLO}
    }\\
    \subfigure[$T_{cc}^{\ast +}\to D^+ D^0\gamma\pi^0  $] {\includegraphics[scale=0.23]{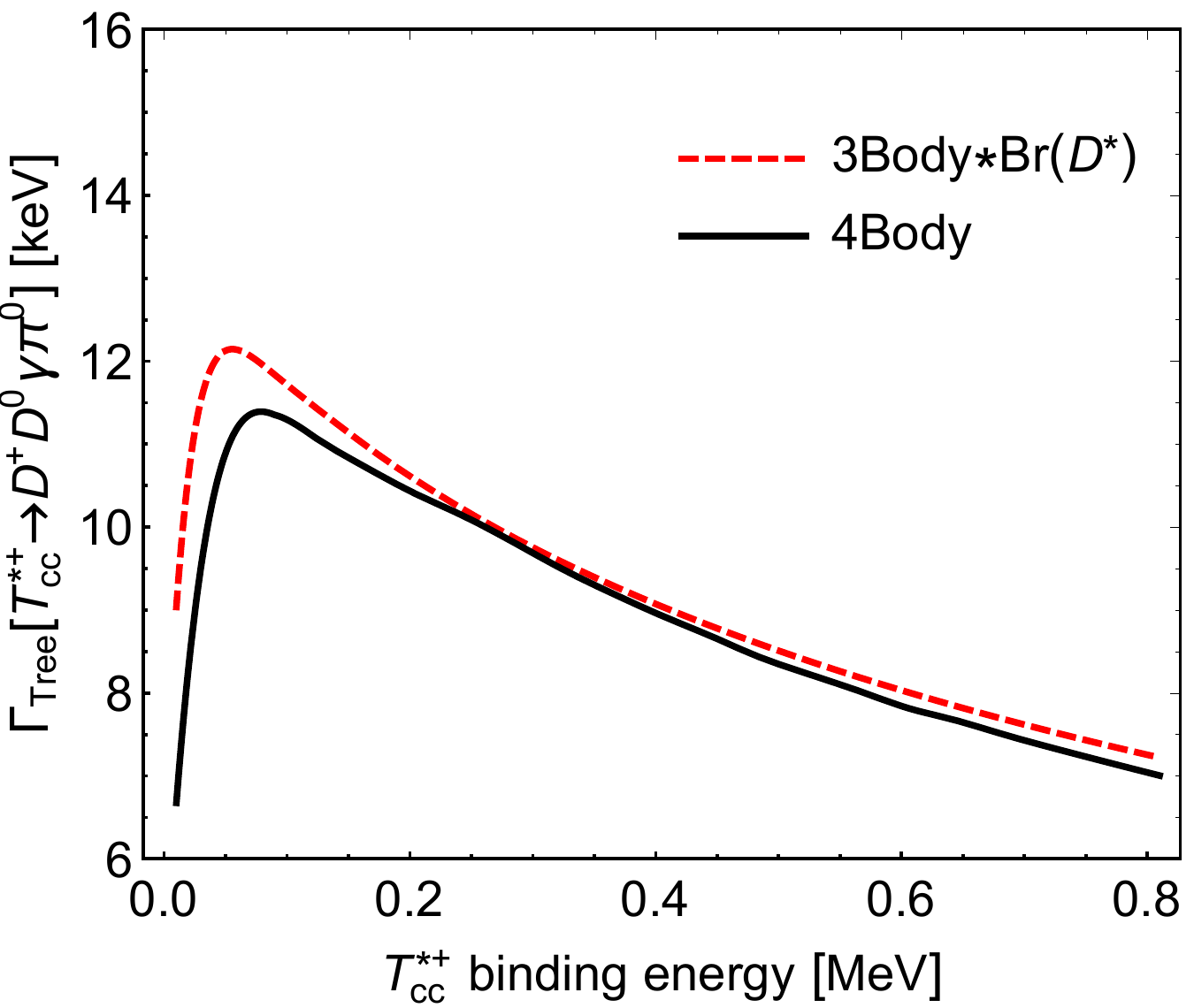} \label{fig_TsDcDpi0g3B4BTree}
    }
    \subfigure[$T_{cc}^{\ast +}\to D^+ D^0 \gamma \pi^0 $] {\includegraphics[scale=0.23]{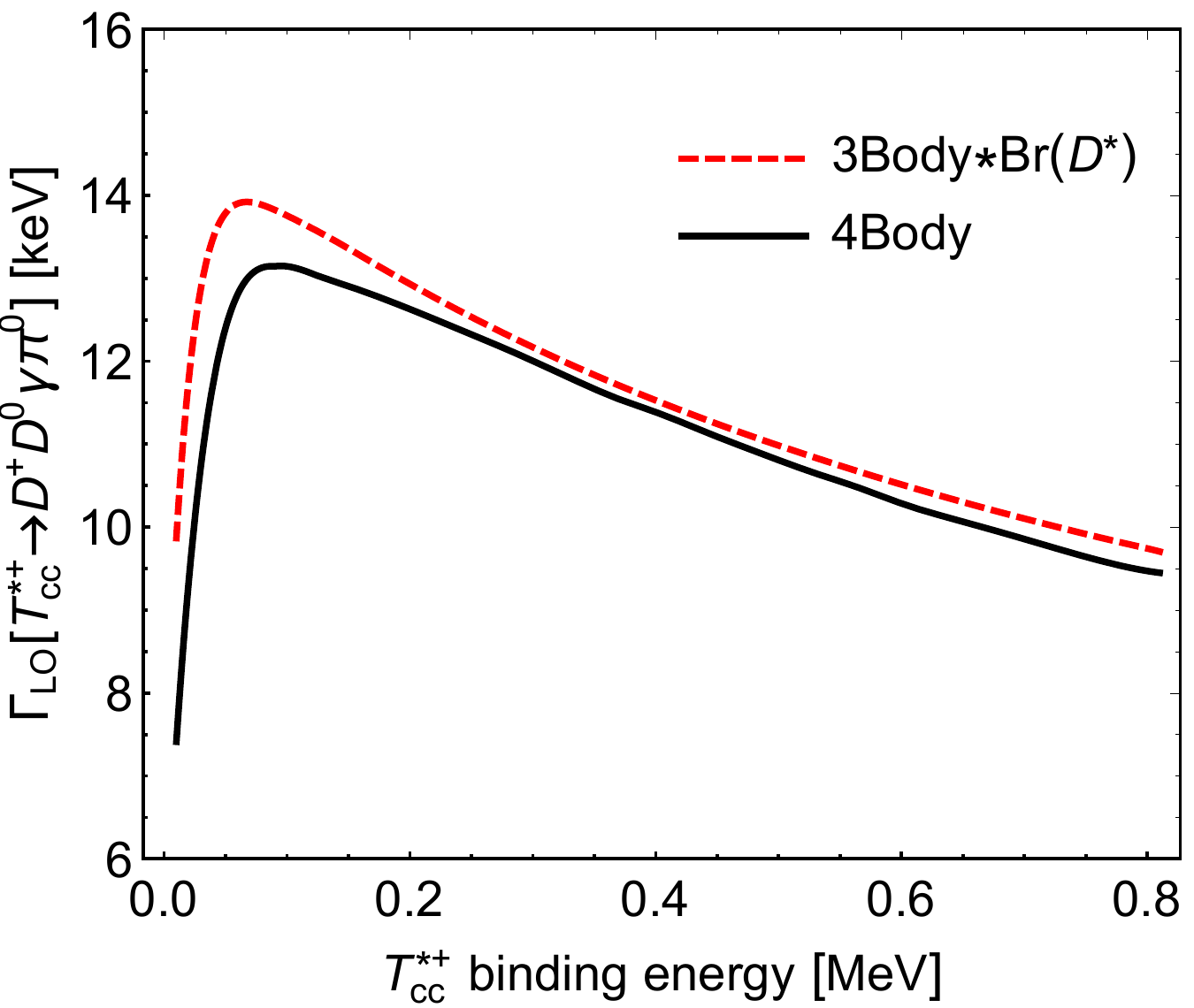} \label{fig_TsDcDpi0g3B4BLO}
    }
    \subfigure[$T_{cc}^{\ast +}\to D^+ D^0 \gamma \pi^0 $] {\includegraphics[scale=0.23]{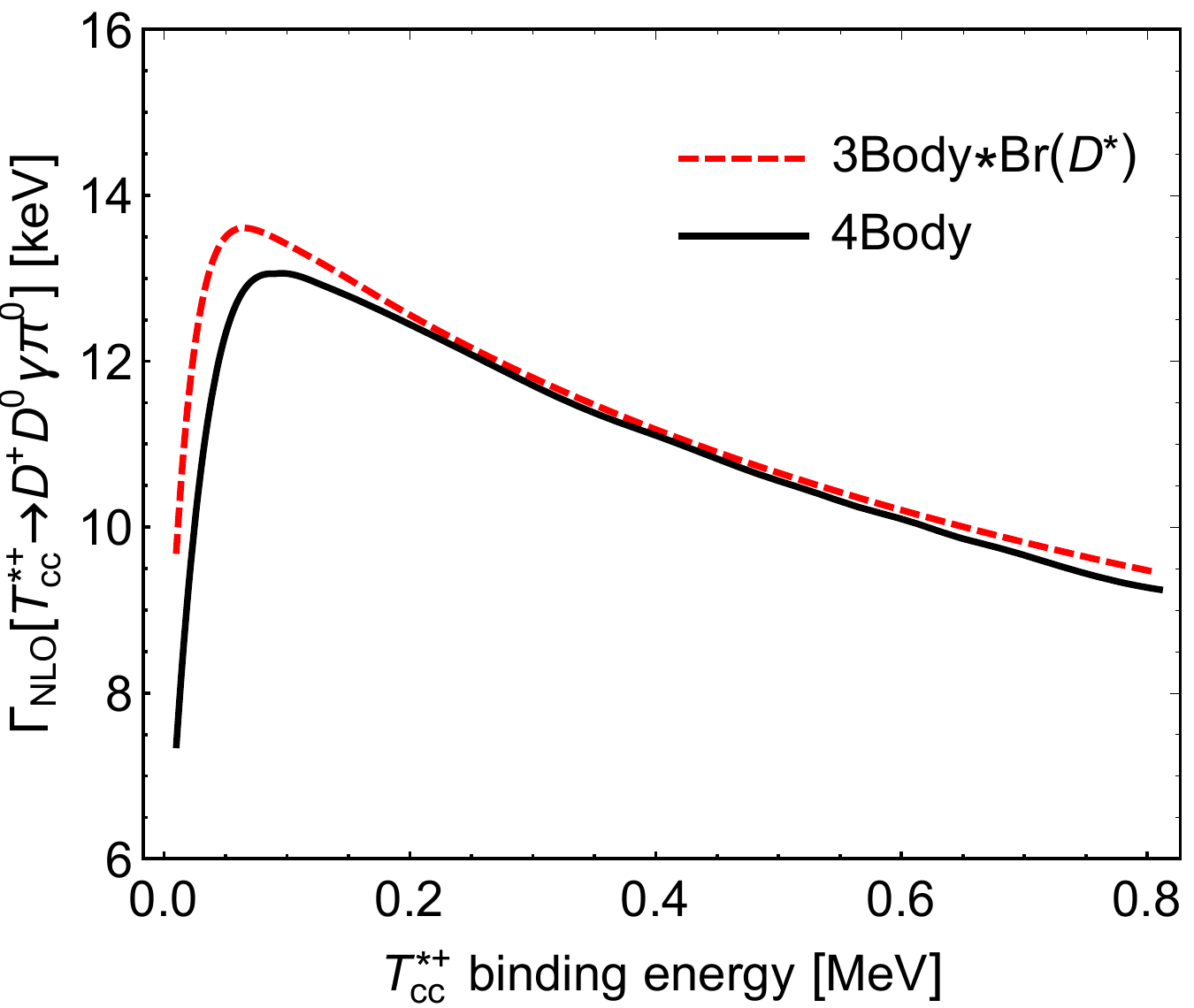} \label{fig_TsDcDpi0g3B4BNLO}
    }
    \subfigure[$T_{cc}^{\ast +}\to D^0 D^0 \gamma \pi^+$] {\includegraphics[scale=0.23]{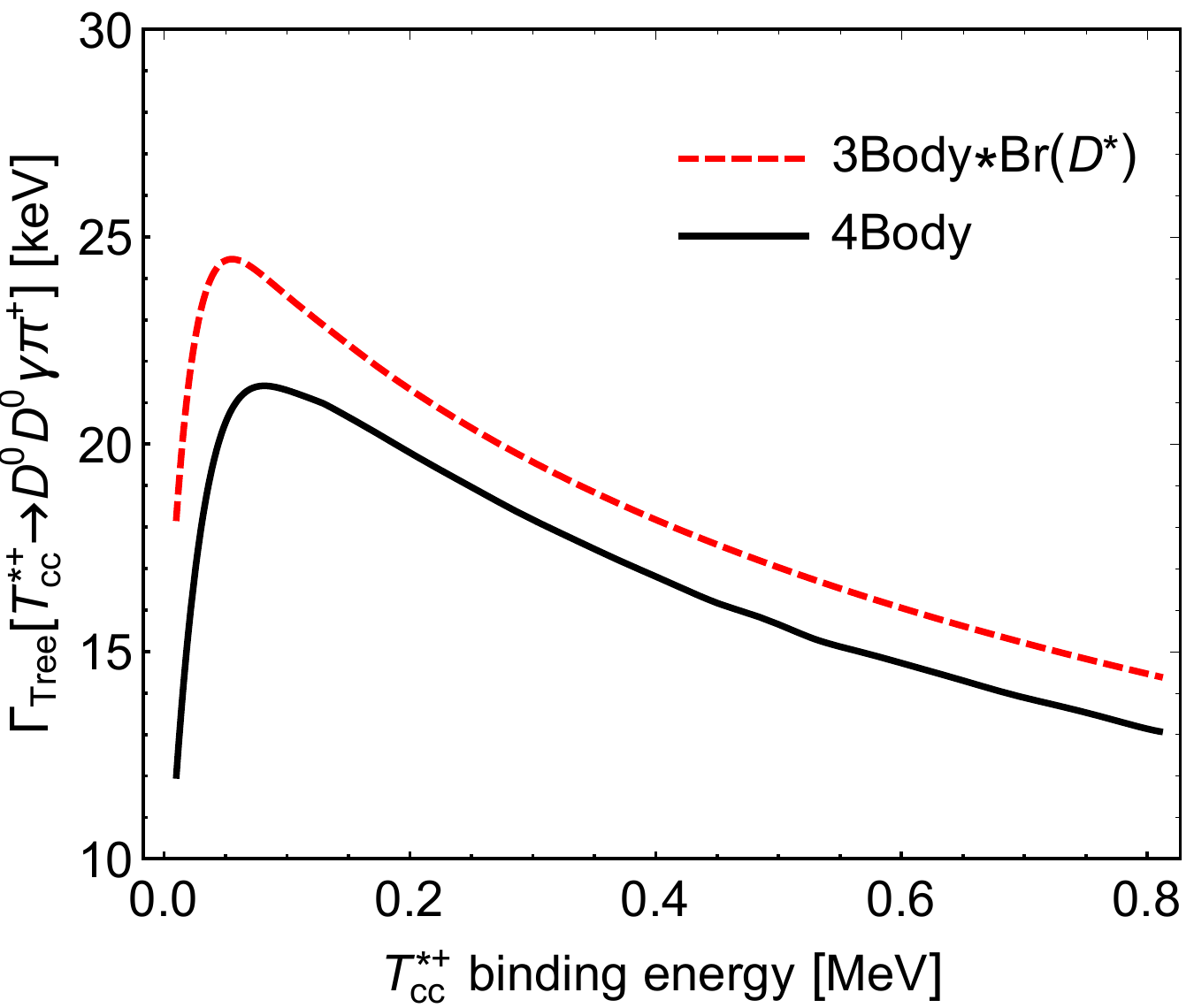} \label{fig_TsDDpicg3B4BTree}
    }
    \subfigure[$T_{cc}^{\ast +}\to D^0 D^0 \gamma \pi^+$] {\includegraphics[scale=0.23]{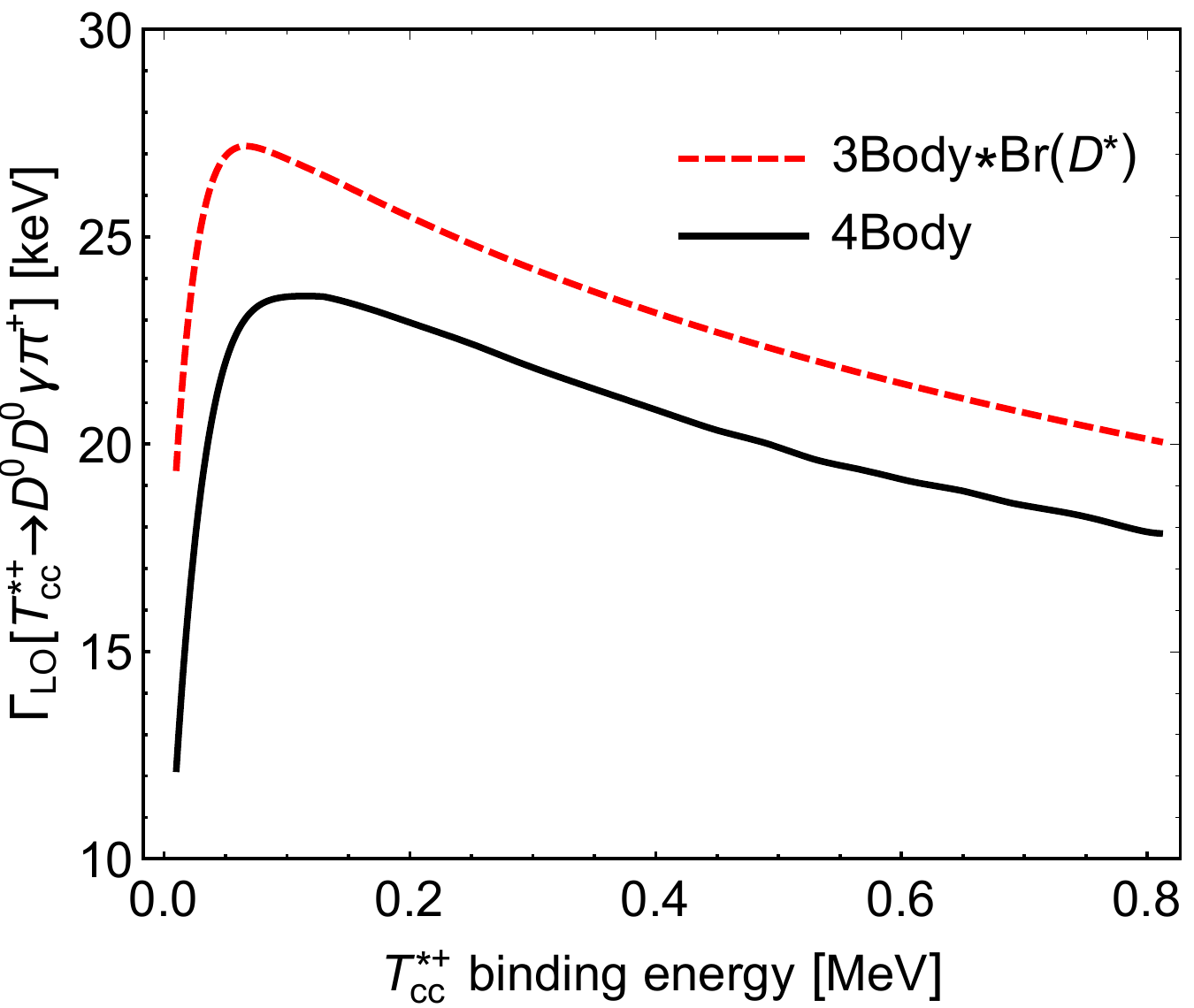} \label{fig_TsDDpicg3B4BLO}
    }
    \subfigure[$T_{cc}^{\ast +}\to D^0 D^0 \gamma \pi^+$] {\includegraphics[scale=0.23]{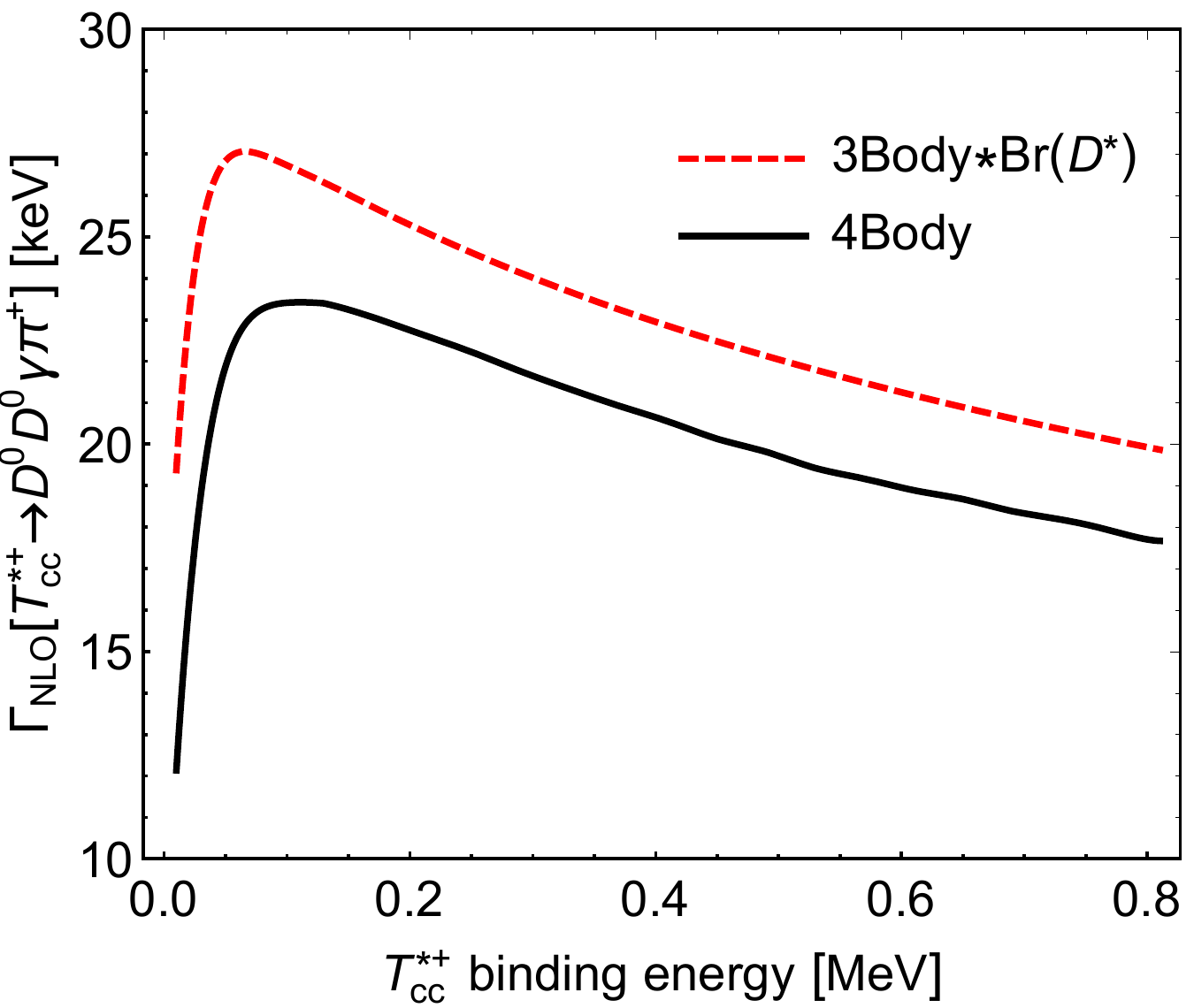} \label{fig_TsDDpicg3B4BNLO}
    }
   \caption{Partial decay widths of the $T_{cc}^{\ast }\rightarrow DD\gamma\gamma$, $DD\gamma\pi$, and $T_{cc}^{*+} \to D^*D \gamma$ or $D^*D\pi$ multiplied by the branch ratio of $D^* \to D\gamma $ or $D\pi$ versus the binding energy of the $T_{cc}^{\ast +}$.}
    \label{fig_TsDDg/pig3B4B}
\end{figure}
\begin{figure}[htbp]
    \subfigure[$T_{cc}^{\ast +}\to D^+ D^0\gamma \gamma $] {\includegraphics[scale=0.23]{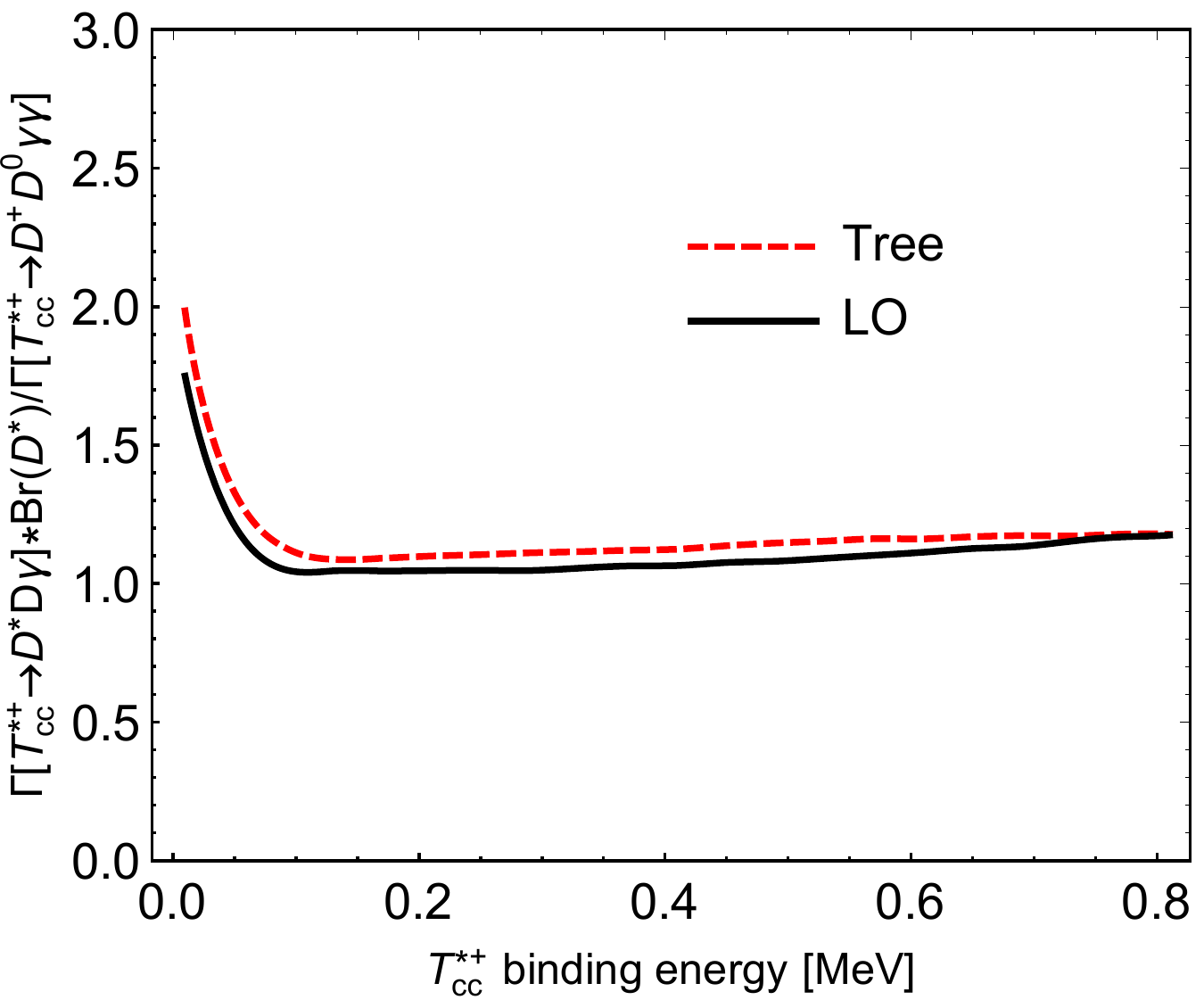} \label{fig_TsDcDgg3Bvs4BTreeLO}
    }
    \subfigure[$T_{cc}^{\ast +}\to D^+ D^0 \gamma\pi^0$] {\includegraphics[scale=0.23]{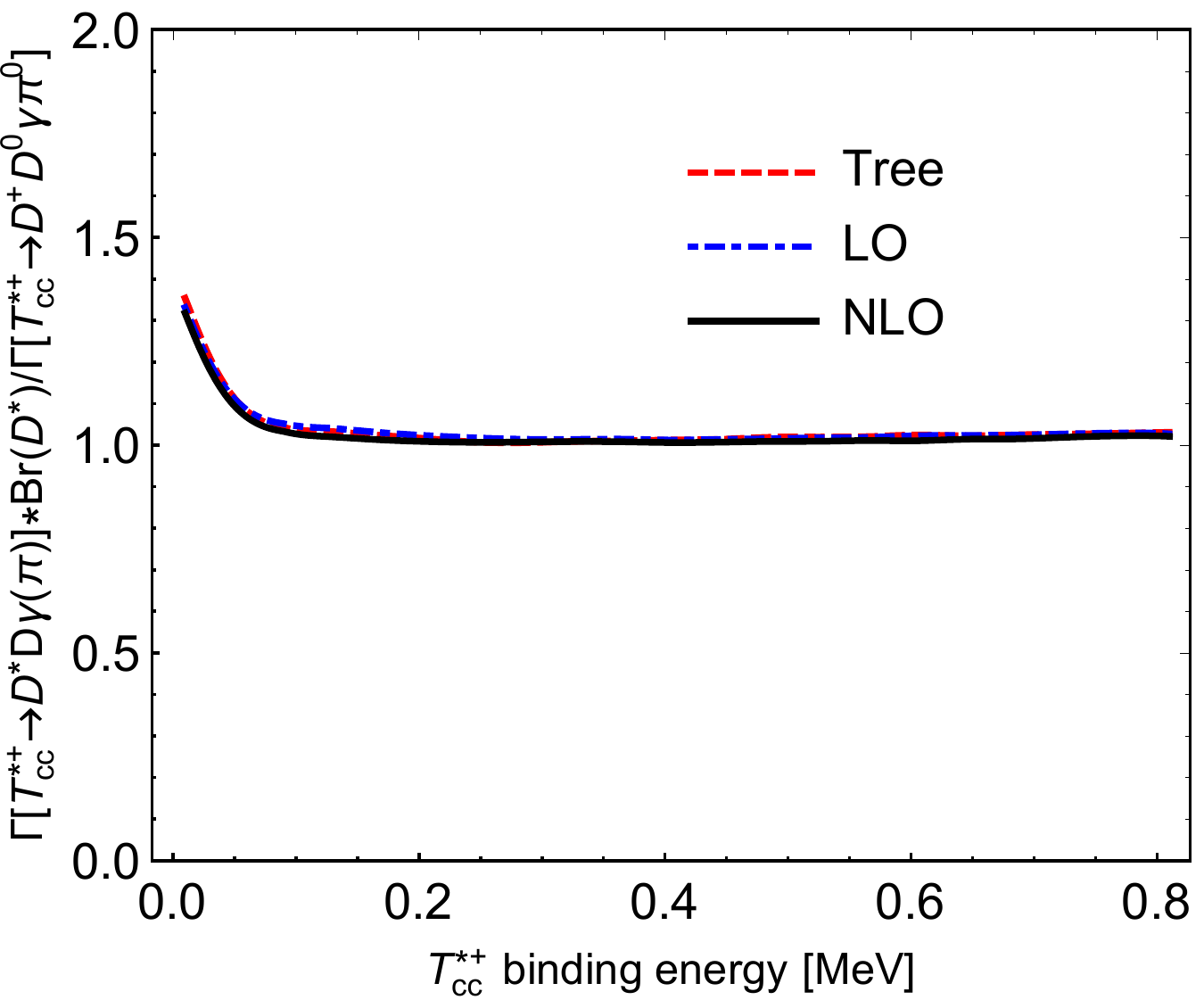} \label{fig_TsDcDpi0g3Bvs4BTreeLONLO}
    }
    \subfigure[$T_{cc}^{\ast +}\to D^0 D^0\gamma\pi^+$] {\includegraphics[scale=0.23]{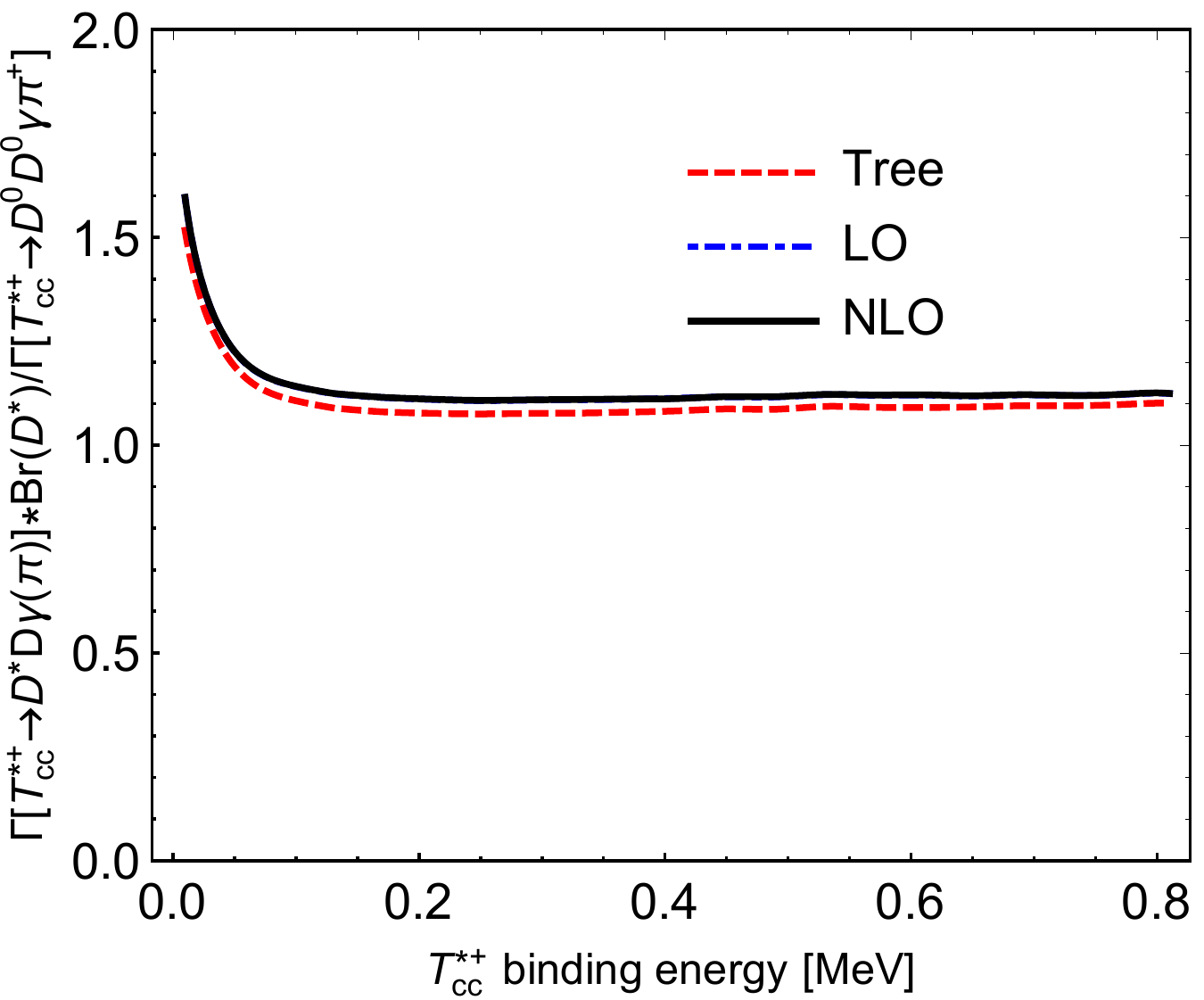} \label{fig_TsDDpicg3Bvs4BTreeLONLO}
    }
   \caption{Ratios between the partial decay widths of the $T_{cc}^{\ast }\rightarrow DD\gamma\gamma(\pi)$ and those obtained using $\Gamma(T_{cc}^{*+} \to D^*D\gamma(\pi)) \times {\rm Br}(D^*\to D \gamma(\pi))$ versus the binding energy of the $T_{cc}^{\ast}$.}
\label{fig_TsDDg/pig3Bvs4B}
\end{figure}
 
Table~\ref{Tab:Tccstar+4Body} shows the radiative decay widths of the $T_{cc}^*$. The second column denoted by $\Gamma_{\text{Tree}}$ includes the contribution from only the tree-level diagram. The third column marked by $\Gamma_{\rm{LO}}$ lists the LO decay widths, including both the tree-level and the $D^*D$ rescattering contribution. The fourth column named NLO lists the results up to NLO including corrections from the $D^*\pi$ rescattering.
We also list the results obtained by multiplying the three-body decay widths into $D^*D\gamma$ or $D^*D\pi$ with the corresponding $D^*\to D\gamma$ or $D^*\to D\pi$ branching fractions, for which the interference between different intermediate three-body decays is ignored. One can see that the difference between the results with and without the interference between different intermediate three-body $D^*D\gamma$ or $D^*D\pi$ channels is marginal. Thus, the $T_{cc}^*$ radiative decay width can be well approximated by summing over the three-body final state $D^*D\gamma$ partial widths, given in Eq.~\eqref{eq:total}.

In Fig.~\ref{fig_TsDDg/pigDW}, we present the partial widths of $T_{cc}^* \to DD\gamma\gamma$ and $T_{cc}^* \to DD\gamma\pi$  varying the binding energy of $T_{cc}^*$ from $0.01$ to $0.80~\rm{MeV}$. To see the relations between the three-body decay $T_{cc}^{*} \rightarrow D^* D \gamma$ and the four-body decay $T_{cc}^{*} \rightarrow DD\gamma\gamma$ or $DD\gamma\pi$ more clearly, we compare the partial decay widths $\Gamma[T_{cc}^{*} \to  DD\gamma\gamma(\pi)]$ and $\Gamma[T_{cc}^{*} \rightarrow D^*D\gamma] \times \mathrm{Br}[D^* \rightarrow D\gamma(\pi)]$ in Fig.~\ref{fig_TsDDg/pig3B4B} and give their ratios in Fig.~\ref{fig_TsDDg/pig3Bvs4B}. One can see that the difference between the decay widths with and without the interference between the intermediate three-body $D^*D\gamma$ states is marginal for the $T_{cc}^*$ binding energy larger than $200 \, \rm{keV}$, and the binding energy $(503 \pm 40)\, \rm{keV}$ predicted in Ref.~\cite{Du:2021zzh} is within this region.

\section{Summary}\label{sec:SUMMARY}

In this work, we calculate the  radiative partial decay widths of  $T_{cc}^{*} \to D^{*+} D^0 \gamma/D^{*0} D^+ \gamma$ taking into account the $D^{\ast} D$ rescattering contributions where the $T_{cc}^{* +}$ is an isoscalar $1^+$ $D^{* +} D^{* 0}$ shallow bound state and the spin partner of the $T_{cc}(3875)$. We found that the $I=0$ $D^{\ast+} D^0/D^{\ast0} D^+$ rescattering, which generates a $T_{cc}^+$ pole just below the threshold, contributes at LO and has a sizable constructive contribution to the partial width of the $T_{cc}^{*+} \to D^{*+} D^0 \gamma$ and destructive influence on the $T_{cc}^{*+} \to D^{* 0} D^{+} \gamma$.
The two-body partial decay widths of the $T_{cc}^{*+} \to T_{cc}^+\gamma$ and $T_{cc}^+\pi^0$ are calculated to be about $6$ and $3~\rm{keV}$, respectively. Since the $D^*$ further decays into the $D\gamma$ and $D\pi$ final states, we also calculate the four-body decay widths of $T_{cc}^{*+} \rightarrow DD\gamma\gamma$ and $DD\gamma\pi$, and find that the interference effect between different intermediate $D^*D\gamma$ and $D^*D\pi$ states is small. Thus, the $T_{cc}^{*}$ radiative decay width can be well approximated by summing over the $D^*D\gamma$ partial widths for the $T_{cc}^*$ binding energy larger than $200\, \rm{keV}$. Taking the binding energy $(503\pm40)\, \rm{keV}$ predicted in Ref.~\cite{Du:2021zzh}, the obtained $T_{cc}^{*}$ radiative decay width is about $24\, \rm{keV}$. Adding the hadronic decay width $41 \, \rm{keV}$ calculated in Ref.~\cite{Jia:2022qwr}, the total width of the $T_{cc}^*$ is about $65\, \rm{keV}$. The results calculated here should be useful for searching the $T_{cc}^{\ast+}$ state at LHCb and testing the molecular nature of the $T_{cc}$ in the future.

\section{Acknowledgments}\label{sec: ACKNOWLEDGMENTS}
This work is supported in part by the Chinese Academy of Sciences under Grant No. XDB34030000; by the National Natural Science Foundation of China (NSFC) under Grants No. 12125507, No. 11835015, No. 12047503, and No. 12075133; and by the NSFC and the Deutsche Forschungsgemeinschaft
(DFG) through the funds provided to the TRR110 “Symmetries and the Emergence of Structure in QCD” (NSFC Grant No. 12070131001, DFG Project-ID No. 196253076). This work is also supported by Taishan Scholar Project of Shandong Province under Grant No. tsqn202103062 and the Higher Educational Youth Innovation Science and Technology
Program Shandong Province  under Grant No. 2020KJJ004.

\appendix

\section{Three-point loop integrals}\label{sec:Triangal loop}

Both the scalar and vector three-point loop integrals are ultraviolet convergent. Their expressions with nonrelativistic propagators are given by~\cite{Guo:2010ak}
\begin{align}
I(q)
&=i\int \frac{d^4l}{(2 \pi)^{4}} \frac{1}{\left(l^0-m_{1}-\frac{\vec{l}^{2}}{2 m_{1}}+i \epsilon\right)\left(M-l^{0}-m_{2}-\frac{\vec{l}^{2}}{2 m_{2}}+i \epsilon\right)\left[l^{0}-q^0-m_{3}-\frac{\left(\vec{l}-\vec{q}\right)^{2}}{2 m_{3}}+i \epsilon\right]}\nonumber \\
&=\int \frac{d^{3}l}{(2 \pi)^{3}} \frac{1}{\left(b_{12}+\frac{\vec{l}^{2}}{2 \mu_{12}}-i \epsilon\right)\left[b_{23}+\frac{\vec{l}^{2}}{2 m_{2}}+\frac{\left(\vec{l}-\vec{q}\right)^{2}}{2 m_{3}}-i \epsilon\right]}\nonumber \\
&=\frac{\mu_{12} \mu_{23}}{2 \pi } \frac{1}{\sqrt{a}}\left[\tan ^{-1}\left(\frac{c_2-c_1}{2 \sqrt{a c_1}}\right)+\tan ^{-1}\left(\frac{2 a+c_1-c_2}{2 \sqrt{a\left(c_2-a\right)}}\right)\right],
\label{Eq:loop_integral} \\
q^{i} I^{(1)}(q)&=\int \frac{d^{3} \vec{l}}{(2 \pi)^{3}}  \frac{l^{i}}{\left(b_{12}+\frac{\vec{l}^{2}}{2 \mu_{12}}-i \epsilon\right)\left[b_{23}+\frac{\vec{l}^{2}}{2 m_{2}}+\frac{\left(\vec{l}-\vec{q}\right)^{2}}{2 m_{3}}-i \epsilon\right]} \nonumber \\
 &=\frac{\mu_{23}}{a m_3}\left[B\left(c_2-a\right)-B\left(c_1\right)+\frac{1}{2}\left(c_2-c_1\right) I(q)\right], \label{Eq.qiqjI02}
\end{align}
where $\mu_{ij}=m_{i} m_{j} /\left(m_{i}+m_{j}\right)$ are the reduced masses, $b_{12}=m_1+m_2-M,b_{23}=m_{2}+m_{3}+q^0-M$, 
\begin{align}
 a=\left(\frac{\mu_{23}}{m_3}\right)^2 \vec{q}^{\, 2}, \quad c_1=2 \mu_{12} b_{12}, \quad c_2=2 \mu_{23} b_{23}+\frac{\mu_{23}}{m_3} \vec{q}^{\,2},
\end{align}
and the two-point function $B(c)$ in the power divergence subtraction (PDS) scheme~\cite{Kaplan:1998tg} reads
\begin{align}
    B(c)&=2 \mu_{12} \mu_{23} \left(\frac{\Lambda_{\mathrm{PDS}}}{2}\right)^{4-d} \int \frac{d^{d-1} l}{(2 \pi)^{d-1}} \frac{1}{\vec{l}^{\,2}+c-i \epsilon} \nonumber \\
&=\frac{\mu_{12} \mu_{23}}{2 \pi}\left(\Lambda_{\mathrm{PDS}}-\sqrt{c-i \epsilon}\right),
\end{align}
with $\Lambda_{\rm{PDS}}$ a scale in the PDS scheme.

The width of the unstable $D^*$ can be included by considering the $D^*$ self-energy contribution shown in Eq.~\eqref{Eq.Dstar_self_energy} by the following replacement:
\begin{align}
    m_{D^*}\to m_{D^*}-i\frac{\Gamma_{D^*} }{2}.
\end{align}

\section{Four-body decay amplitudes\label{Appendix:4body_amplitudes}}

In this section, we show all the amplitudes for the diagrams in Figs.~\ref{fig_TsDcDggFD}-\ref{fig_TsDDpicgFD} of the four-body $T_{cc}^{*} \to DD\gamma\gamma$ and $DD\gamma\pi$ decays. 

\subsection{$T_{cc}^{*+}\to D^{+}D^0\gamma\gamma$ amplitudes}

We first consider the decay $T_{cc}^{*+} \to D^{+}D^0\gamma\gamma$. The LO amplitude from the tree diagram in Fig.~\ref{fig_TsDcDggFDa} reads
\begin{align}
&i \mathcal{A}_{\text{a}}[T_{cc}^{*+} \to  \gamma(p_1)D^0(p_2)\gamma(p_3)D^+(p_4)]=\frac{-i g_s \mu_{D^0}\mu_{D^+}}{4\sqrt{2}}\frac{1}{q^0-m_{D^{*0}}-\frac{\vec{q}^{\, 2}}{2m_{D^{*0}}}+i\frac{\Gamma_{D^{*0}}}{2}}\nonumber\\
&\, \times \frac{1}{k^0-m_{D^{*+}}-\frac{\vec{k}^2}{2m_{D^{*+}}}+i\frac{\Gamma_{D^{*+}}}{2}}\epsilon^{ijk}\epsilon^i(T_{cc}^{*+})\epsilon^{kmn}p_1^m\epsilon^{n*}(\gamma_1)\epsilon^{jst}p_3^s\epsilon^{t*}(\gamma_3),
\end{align}
with $s_{12}=q^2$, $s_{34}=k^2$, and $s_{ij}=(p_i+p_j)^2$, $i,j=1,...,4$. Here, $p_1^j$ and $p_3^k$ are the three-momenta of the two photons in the final state in the $T_{cc}^{*}$ rest frame, respectively, and $q^{\mu}=(q^0, \vec{q})$, $k^{\mu}=(k^0, \vec{k})$ are the four-momenta of the $(1, 2)$ and $(3, 4)$ two-particle systems in the $T_{cc}^{*}$ rest frame, respectively.

Considering the crossed-channel effects of the two identical photons in the final state, we also have
\begin{align}
&i \mathcal{A}_{\text{a}}[T_{cc}^{*+} \to  \gamma(p_3)D^0(p_2)\gamma(p_1)D^+(p_4)]=\frac{i g_s \mu_{D^0}\mu_{D^+}}{4\sqrt{2}}\frac{1}{l^0-m_{D^{*0}}-\frac{\vec{l}^2}{2m_{D^{*0}}}+i\frac{\Gamma_{D^{*0}}}{2}}\nonumber\\
&\, \times \frac{1}{t^0-m_{D^{*+}}-\frac{\vec{t}^2}{2m_{D^{*+}}}+i\frac{\Gamma_{D^{*+}}}{2}}\epsilon^{ijk}\epsilon^i(T_{cc}^{*+})\epsilon^{kmn}p_1^m\epsilon^{n*}(\gamma_1)\epsilon^{jst}p_3^s\epsilon^{t*}(\gamma_3),
\end{align}
where $s_{23}=l^2$, $s_{14}=t^2$, and $l^{\mu}=(l^0, \vec{l})$, $t^{\mu}=(t^0, \vec{t})$ are the four-momenta of the $(2, 3)$ and $(1, 4)$ two-particle systems in the $T_{cc}^{*}$ rest frame, respectively.

The LO amplitudes from the $D^{*+}D^0/D^{*0}D^+$ rescattering diagrams in Figs.~\ref{fig_TsDcDggFDb}-\ref{fig_TsDcDggFDe} read
\begin{align}
&i \mathcal{A}_{\text{b}}[T_{cc}^{*+} \to  \gamma(p_1)D^0(p_2)\gamma(p_3)D^+(p_4)]=\frac{i g_s \mu_{D^0}\mu_{D^+} C_{0D1}}{4\sqrt{2}}\frac{1}{(p_3^0+p_4^0)-m_{D^{*+}}-\frac{(\vec{p}_{3}+\vec{p}_{4})^2}{2m_{D^{*+}}}+i\frac{\Gamma_{D^{*+}}}{2}}\nonumber\\
&\, \times \epsilon^{ijk}\epsilon^i(T_{cc}^{*+})\epsilon^{kmn}p_1^m\epsilon^{n*}(\gamma_1)\epsilon^{jst}p_3^s\epsilon^{t*}(\gamma_3)I(p_1),
\end{align}
\begin{align}
&i \mathcal{A}_{\text{b}}[T_{cc}^{*+} \to  \gamma(p_3)D^0(p_2)\gamma(p_1)D^+(p_4)]=\frac{-i g_s \mu_{D^0}\mu_{D^+} C_{0D1}}{4\sqrt{2}}\frac{1}{(p_1^0+p_4^0)-m_{D^{*+}}-\frac{(\vec{p}_{1}+\vec{p}_{4})^2}{2m_{D^{*+}}}+i\frac{\Gamma_{D^{*+}}}{2}}\nonumber\\
&\, \times \epsilon^{ijk}\epsilon^i(T_{cc}^{*+})\epsilon^{kmn}p_1^m\epsilon^{n*}(\gamma_1)\epsilon^{jst}p_3^s\epsilon^{t*}(\gamma_3)I(p_3),
\end{align}
\begin{align}
&i \mathcal{A}_{\text{c}}[T_{cc}^{*+} \to  \gamma(p_1)D^0(p_2)\gamma(p_3)D^+(p_4)]=\frac{-i g_s \mu_{D^+}\mu_{D^+} C_{0D1ex}}{4\sqrt{2}}\frac{1}{(p_3^0+p_4^0)-m_{D^{*+}}-\frac{(\vec{p}_{3}+\vec{p}_{4})^2}{2m_{D^{*+}}}+i\frac{\Gamma_{D^{*+}}}{2}}\nonumber\\
&\, \times \epsilon^{ijk}\epsilon^i(T_{cc}^{*+})\epsilon^{kmn}p_1^m\epsilon^{n*}(\gamma_1)\epsilon^{jst}p_3^s\epsilon^{t*}(\gamma_3)I(p_1),
\end{align}
\begin{align}
&i \mathcal{A}_{\text{c}}[T_{cc}^{*+} \to  \gamma(p_3)D^0(p_2)\gamma(p_1)D^+(p_4)]=\frac{i g_s \mu_{D^+}\mu_{D^+} C_{0D1ex}}{4\sqrt{2}}\frac{1}{(p_1^0+p_4^0)-m_{D^{*+}}-\frac{(\vec{p}_{1}+\vec{p}_{4})^2}{2m_{D^{*+}}}+i\frac{\Gamma_{D^{*+}}}{2}}\nonumber\\
&\, \times \epsilon^{ijk}\epsilon^i(T_{cc}^{*+})\epsilon^{kmn}p_1^m\epsilon^{n*}(\gamma_1)\epsilon^{jst}p_3^s\epsilon^{t*}(\gamma_3)I(p_3),
\end{align}
\begin{align}
&i \mathcal{A}_{\text{d}}[T_{cc}^{*+} \to  \gamma(p_1)D^0(p_2)\gamma(p_3)D^+(p_4)]=\frac{i g_s \mu_{D^0}\mu_{D^+} C_{0D2}}{4\sqrt{2}}\frac{1}{(p_1^0+p_2^0)-m_{D^{*0}}-\frac{(\vec{p}_{1}+\vec{p}_{2})^2}{2m_{D^{*0}}}+i\frac{\Gamma_{D^{*0}}}{2}}\nonumber\\
&\, \times \epsilon^{ijk}\epsilon^i(T_{cc}^{*+})\epsilon^{kmn}p_1^m\epsilon^{n*}(\gamma_1)\epsilon^{jst}p_3^s\epsilon^{t*}(\gamma_3)I(p_3),
\end{align}
\begin{align}
&i \mathcal{A}_{\text{d}}[T_{cc}^{*+} \to  \gamma(p_3)D^0(p_2)\gamma(p_1)D^+(p_4)]=\frac{-i g_s \mu_{D^0}\mu_{D^+} C_{0D2}}{4\sqrt{2}}\frac{1}{(p_3^0+p_2^0)-m_{D^{*0}}-\frac{(\vec{p}_{3}+\vec{p}_{2})^2}{2m_{D^{*0}}}+i\frac{\Gamma_{D^{*0}}}{2}}\nonumber\\
&\, \times \epsilon^{ijk}\epsilon^i(T_{cc}^{*+})\epsilon^{kmn}p_1^m\epsilon^{n*}(\gamma_1)\epsilon^{jst}p_3^s\epsilon^{t*}(\gamma_3)I(p_1),
\end{align}
\begin{align}
&i \mathcal{A}_{\text{e}}[T_{cc}^{*+} \to  \gamma(p_1)D^0(p_2)\gamma(p_3)D^+(p_4)]=\frac{-i g_s \mu_{D^0}\mu_{D^0} C_{0D2ex}}{4\sqrt{2}}\frac{1}{(p_1^0+p_2^0)-m_{D^{*0}}-\frac{(\vec{p}_{1}+\vec{p}_{2})^2}{2m_{D^{*0}}}+i\frac{\Gamma_{D^{*0}}}{2}}\nonumber\\
&\, \times \epsilon^{ijk}\epsilon^i(T_{cc}^{*+})\epsilon^{kmn}p_1^m\epsilon^{n*}(\gamma_1)\epsilon^{jst}p_3^s\epsilon^{t*}(\gamma_3)I(p_3),
\end{align}
\begin{align}
&i \mathcal{A}_{\text{e}}[T_{cc}^{*+} \to  \gamma(p_3)D^0(p_2)\gamma(p_1)D^+(p_4)]=\frac{i g_s \mu_{D^0}\mu_{D^0} C_{0D2ex}}{4\sqrt{2}}\frac{1}{(p_3^0+p_2^0)-m_{D^{*0}}-\frac{(\vec{p}_{3}+\vec{p}_{2})^2}{2m_{D^{*0}}}+i\frac{\Gamma_{D^{*0}}}{2}}\nonumber\\
&\, \times \epsilon^{ijk}\epsilon^i(T_{cc}^{*+})\epsilon^{kmn}p_1^m\epsilon^{n*}(\gamma_1)\epsilon^{jst}p_3^s\epsilon^{t*}(\gamma_3)I(p_1),
\end{align}
 where $p_1^{\mu}=(p_1^0, \vec{p}_1)$, $p_2^{\mu}=(p_2^0, \vec{p}_2)$, $p_3^{\mu}=(p_3^0, \vec{p}_3)$, and $p_4^{\mu}=(p_4^0, \vec{p}_4)$ are the four-momenta of the four final-state particles in the rest frame of the $T_{cc}^{*}$, respectively. 
The masses $m_1$, $m_2$, and $m_3$ in the loop integrals $I(p_i)$ are taken to be the masses of $D^{*0}$, $D^{*+}$, and $D^0$ in Figs.~\ref{fig_TsDcDggFDb} and \ref{fig_TsDcDggFDe} and the masses of $D^{*+}$, $D^{*0}$, and $D^+$ in Figs.~\ref{fig_TsDcDggFDc} and \ref{fig_TsDcDggFDd}, respectively.

\subsection{$T_{cc}^{*+}\to D^{+}D^0\gamma\pi^0$ amplitudes}

For the decay $T_{cc}^{*+} \to D^{+} D^0\gamma\pi^0$, the LO amplitude from the tree diagram in Fig.~\ref{fig_TsDcDpi0gFDa} reads
\begin{align}
&i \mathcal{A}_{\text{a}}[T_{cc}^{*+} \to  \gamma(p_1)D^0(p_2)\pi^0(p_3)D^+(p_4)]=\frac{i g_s \bar{g} \mu_{D^0}}{2\sqrt{2}F_{\pi}\sqrt{2m_{\pi^0}}}\frac{1}{q^0-m_{D^{*0}}-\frac{\vec{q}^2}{2m_{D^{*0}}}+i\frac{\Gamma_{D^{*0}}}{2}}\nonumber\\
&\, \times \frac{1}{k^0-m_{D^{*+}}-\frac{\vec{k}^2}{2m_{D^{*+}}}+i\frac{\Gamma_{D^{*+}}}{2}}\epsilon^{ijk}\epsilon^i(T_{cc}^{*+})p_3^j\epsilon^{kmn}p_1^m\epsilon^{n*}(\gamma),
\end{align}
\begin{align}
&i \mathcal{A}_{\text{b}}[T_{cc}^{*+} \to  \gamma(p_1)D^0(p_2)\pi^0(p_3)D^+(p_4)]=\frac{i g_s \bar{g} \mu_{D^+}}{2\sqrt{2}F_{\pi}\sqrt{2m_{\pi^0}}}\frac{1}{t^0-m_{D^{*+}}-\frac{\vec{t}^2}{2m_{D^{*+}}}+i\frac{\Gamma_{D^{*+}}}{2}}\nonumber\\
&\, \times \frac{1}{l^0-m_{D^{*0}}-\frac{\vec{l}^2}{2m_{D^{*0}}}+i\frac{\Gamma_{D^{*0}}}{2}}\epsilon^{ijk}\epsilon^i(T_{cc}^{*+})p_3^j\epsilon^{kmn}p_1^m\epsilon^{n*}(\gamma).
\end{align}

The LO amplitudes from the $D^{*+}D^0/D^{*0}D^+$ rescattering diagrams are
\begin{align}
&i \mathcal{A}_{\text{c}}[T_{cc}^{*+} \to  \gamma(p_1)D^0(p_2)\pi^0(p_3)D^+(p_4)]=\frac{-i g_s \bar{g} \mu_{D^0} C_{0D1}}{2\sqrt{2} F_{\pi} \sqrt{2m_{\pi^0}}}\frac{1}{(p_3^0+p_4^0)-m_{D^{*+}}-\frac{(\vec{p}_{3}+\vec{p}_{4})^2}{2m_{D^{*+}}}+i\frac{\Gamma_{D^{*+}}}{2}}\nonumber\\
&\, \times \epsilon^{ijk}\epsilon^i(T_{cc}^{*+}) p_3^j \epsilon^{kmn}p_1^m\epsilon^{n*}(\gamma)I(p_1),
\end{align}
\begin{align}
&i \mathcal{A}_{\text{d}}[T_{cc}^{*+} \to  \gamma(p_1)D^0(p_2)\pi^0(p_3)D^+(p_4)]=\frac{i g_s \bar{g} \mu_{D^+} C_{0D1ex}}{2\sqrt{2} F_{\pi} \sqrt{2m_{\pi^0}}}\frac{1}{(p_3^0+p_4^0)-m_{D^{*+}}-\frac{(\vec{p}_{3}+\vec{p}_{4})^2}{2m_{D^{*+}}}+i\frac{\Gamma_{D^{*+}}}{2}}\nonumber\\
&\, \times \epsilon^{ijk}\epsilon^i(T_{cc}^{*+}) p_3^j \epsilon^{kmn}p_1^m\epsilon^{n*}(\gamma)I(p_1),
\end{align}
\begin{align}
&i \mathcal{A}_{\text{e}}[T_{cc}^{*+} \to  \gamma(p_1)D^0(p_2)\pi^0(p_3)D^+(p_4)]=\frac{-i g_s \bar{g} \mu_{D^+} C_{0D2}}{2\sqrt{2} F_{\pi} \sqrt{2m_{\pi^0}}}\frac{1}{(p_2^0+p_3^0)-m_{D^{*0}}-\frac{(\vec{p}_{2}+\vec{p}_{3})^2}{2m_{D^{*0}}}+i\frac{\Gamma_{D^{*0}}}{2}}\nonumber\\
&\, \times \epsilon^{ijk}\epsilon^i(T_{cc}^{*+}) p_3^j \epsilon^{kmn}p_1^m\epsilon^{n*}(\gamma)I(p_1),
\end{align}
\begin{align}
&i \mathcal{A}_{\text{f}}[T_{cc}^{*+} \to  \gamma(p_1)D^0(p_2)\pi^0(p_3)D^+(p_4)]=\frac{i g_s \bar{g} \mu_{D^0} C_{0D2ex}}{2\sqrt{2} F_{\pi} \sqrt{2m_{\pi^0}}}\frac{1}{(p_2^0+p_3^0)-m_{D^{*0}}-\frac{(\vec{p}_{2}+\vec{p}_{3})^2}{2m_{D^{*0}}}+i\frac{\Gamma_{D^{*0}}}{2}}\nonumber\\
&\, \times \epsilon^{ijk}\epsilon^i(T_{cc}^{*+}) p_3^j \epsilon^{kmn}p_1^m\epsilon^{n*}(\gamma)I(p_1),
\end{align}
\begin{align}
&i \mathcal{A}_{\text{g}}[T_{cc}^{*+} \to  \gamma(p_1)D^0(p_2)\pi^0(p_3)D^+(p_4)]=\frac{-i g_s \bar{g} \mu_{D^+} C_{0D1}}{2\sqrt{2} F_{\pi} \sqrt{2m_{\pi^0}}}\frac{1}{(p_1^0+p_4^0)-m_{D^{*+}}-\frac{(\vec{p}_{1}+\vec{p}_{4})^2}{2m_{D^{*+}}}+i\frac{\Gamma_{D^{*+}}}{2}}\nonumber\\
&\, \times \epsilon^{ijk}\epsilon^i(T_{cc}^{*+}) p_3^j \epsilon^{kmn}p_1^m\epsilon^{n*}(\gamma)I(p_3),
\end{align}
\begin{align}
&i \mathcal{A}_{\text{h}}[T_{cc}^{*+} \to  \gamma(p_1)D^0(p_2)\pi^0(p_3)D^+(p_4)]=\frac{-i g_s \bar{g} \mu_{D^+} C_{0D1ex}}{2\sqrt{2} F_{\pi} \sqrt{2m_{\pi^0}}}\frac{1}{(p_1^0+p_4^0)-m_{D^{*+}}-\frac{(\vec{p}_{1}+\vec{p}_{4})^2}{2m_{D^{*+}}}+i\frac{\Gamma_{D^{*+}}}{2}}\nonumber\\
&\, \times \epsilon^{ijk}\epsilon^i(T_{cc}^{*+}) p_3^j \epsilon^{kmn}p_1^m\epsilon^{n*}(\gamma)I(p_3),
\end{align}
\begin{align}
&i \mathcal{A}_{\text{i}}[T_{cc}^{*+} \to  \gamma(p_1)D^0(p_2)\pi^0(p_3)D^+(p_4)]=\frac{-i g_s \bar{g} \mu_{D^0} C_{0D2}}{2\sqrt{2} F_{\pi} \sqrt{2m_{\pi^0}}}\frac{1}{(p_1^0+p_2^0)-m_{D^{*0}}-\frac{(\vec{p}_{1}+\vec{p}_{2})^2}{2m_{D^{*0}}}+i\frac{\Gamma_{D^{*0}}}{2}}\nonumber\\
&\, \times \epsilon^{ijk}\epsilon^i(T_{cc}^{*+}) p_3^j \epsilon^{kmn}p_1^m\epsilon^{n*}(\gamma)I(p_3),
\end{align}
\begin{align}
&i \mathcal{A}_{\text{j}}[T_{cc}^{*+} \to  \gamma(p_1)D^0(p_2)\pi^0(p_3)D^+(p_4)]=\frac{-i g_s \bar{g} \mu_{D^0} C_{0D2ex}}{2\sqrt{2} F_{\pi} \sqrt{2m_{\pi^0}}}\frac{1}{(p_1^0+p_2^0)-m_{D^{*0}}-\frac{(\vec{p}_{1}+\vec{p}_{2})^2}{2m_{D^{*0}}}+i\frac{\Gamma_{D^{*0}}}{2}}\nonumber\\
&\, \times \epsilon^{ijk}\epsilon^i(T_{cc}^{*+}) p_3^j \epsilon^{kmn}p_1^m\epsilon^{n*}(\gamma)I(p_3),
\end{align}
where the masses $m_1$, $m_2$, and $m_3$ in the loop integrals $I(p_i)$ are taken to be the masses of $D^{*0}$, $D^{*+}$, and $D^0$ in Figs.~\ref{fig_TsDcDpi0gFDc}, \ref{fig_TsDcDpi0gFDf}, \ref{fig_TsDcDpi0gFDg} and \ref{fig_TsDcDpi0gFDj}, and the masses of $D^{*+}$, $D^{*0}$, and $D^+$ in Figs.~\ref{fig_TsDcDpi0gFDd}, \ref{fig_TsDcDpi0gFDe}, \ref{fig_TsDcDpi0gFDh} and \ref{fig_TsDcDpi0gFDi}, respectively.

The NLO amplitudes from the $D^*\pi$ rescattering diagrams in Figs.~\ref{fig_TsDcDpi0gFDk}-\ref{fig_TsDcDpi0gFDn} are
\begin{align}
&i \mathcal{A}_{\text{k}}[T_{cc}^{*+} \to  \gamma(p_1)D^0(p_2)\pi^0(p_3)D^+(p_4)]=\frac{-i g_s \bar{g} \mu_{D^+} C_{\pi1}}{4\sqrt{2} m_{\pi^0} F_{\pi} \sqrt{2m_{\pi^0}}}\frac{1}{(p_1^0+p_4^0)-m_{D^{*+}}-\frac{(\vec{p}_{1}+\vec{p}_{4})^2}{2m_{D^{*+}}}+i\frac{\Gamma_{D^{*+}}}{2}}\nonumber\\
&\, \times \epsilon^{ijk}\epsilon^i(T_{cc}^{*+}) p_2^j \epsilon^{kmn}p_1^m\epsilon^{n*}(\gamma)\left [I^{(1)}(p_2)-I(p_2)\right],
\end{align}
\begin{align}
&i \mathcal{A}_{\text{l}}[T_{cc}^{*+} \to  \gamma(p_1)D^0(p_2)\pi^0(p_3)D^+(p_4)]=\frac{-i g_s \sqrt{2} \bar{g} \mu_{D^+} C_{\pi1ex}}{4\sqrt{2} \sqrt{m_{\pi^0}m_{\pi^+}} F_{\pi} \sqrt{2m_{\pi^+}}}\frac{1}{(p_1^0+p_4^0)-m_{D^{*+}}-\frac{(\vec{p}_{1}+\vec{p}_{4})^2}{2m_{D^{*+}}}+i\frac{\Gamma_{D^{*+}}}{2}}\nonumber\\
&\, \times \epsilon^{ijk}\epsilon^i(T_{cc}^{*+}) p_2^j \epsilon^{kmn}p_1^m\epsilon^{n*}(\gamma)\left [I^{(1)}(p_2)+I(p_2)\right],
\end{align}
\begin{align}
&i \mathcal{A}_{\text{m}}[T_{cc}^{*+} \to  \gamma(p_1)D^0(p_2)\pi^0(p_3)D^+(p_4)]=\frac{i g_s \bar{g} \mu_{D^0} C_{\pi2}}{4\sqrt{2} m_{\pi^0} F_{\pi} \sqrt{2m_{\pi^0}}}\frac{1}{(p_1^0+p_2^0)-m_{D^{*0}}-\frac{(\vec{p}_{1}+\vec{p}_{2})^2}{2m_{D^{*0}}}+i\frac{\Gamma_{D^{*0}}}{2}}\nonumber\\
&\, \times \epsilon^{ijk}\epsilon^i(T_{cc}^{*+}) p_4^j \epsilon^{kmn}p_1^m\epsilon^{n*}(\gamma)\left [I^{(1)}(p_4)+I(p_4)\right],
\end{align}
\begin{align}
&i \mathcal{A}_{\text{n}}[T_{cc}^{*+} \to  \gamma(p_1)D^0(p_2)\pi^0(p_3)D^+(p_4)]=\frac{-i g_s \sqrt{2} \bar{g} \mu_{D^0} C_{\pi2ex}}{4\sqrt{2} \sqrt{m_{\pi^0} m_{\pi^-}} F_{\pi} \sqrt{2m_{\pi^-}}}\frac{1}{(p_1^0+p_2^0)-m_{D^{*0}}-\frac{(\vec{p}_{1}+\vec{p}_{2})^2}{2m_{D^{*0}}}+i\frac{\Gamma_{D^{*0}}}{2}}\nonumber\\
&\, \times \epsilon^{ijk}\epsilon^i(T_{cc}^{*+}) p_4^j \epsilon^{kmn}p_1^m\epsilon^{n*}(\gamma)\left [I^{(1)}(p_4)-I(p_4)\right],
\end{align}
where the masses $m_1$, $m_2$, and $m_3$ in the loop integrals $I^{(1)}(p_i)$ or $I(p_i)$ are taken to be the masses of $D^{*0}$, $D^{*+}$, and $\pi^0$ in Fig.~\ref{fig_TsDcDpi0gFDk}, the masses of $D^{*+}$, $D^{*0}$, and $\pi^+$ in Fig.~\ref{fig_TsDcDpi0gFDl}, the masses of $D^{*+}$, $D^{*0}$, and $\pi^0$ in Fig.~\ref{fig_TsDcDpi0gFDm}, and the masses of $D^{*0}$, $D^{*+}$, and $\pi^-$ in Fig.~\ref{fig_TsDcDpi0gFDn}, respectively.

\subsection{$T_{cc}^{*+}\to D^0D^0\gamma\pi^+$ amplitudes}

For the decay $T_{cc}^{*+} \to D^0 D^0 \gamma \pi^+$, the LO amplitude from the tree diagram in Fig.~\ref{fig_TsDDpicgFDa} reads
\begin{align}
&i \mathcal{A}_{\text{a}}[T_{cc}^{*+} \to  \gamma(p_1)D^0(p_2)\pi^+(p_3)D^0(p_4)]=\frac{-i g_s \sqrt{2} \bar{g} \mu_{D^0}}{2\sqrt{2}F_{\pi}\sqrt{2m_{\pi^+}}}\frac{1}{q^0-m_{D^{*0}}-\frac{\vec{q}^2}{2m_{D^{*0}}}+i\frac{\Gamma_{D^{*0}}}{2}}\nonumber\\
&\, \times \frac{1}{k^0-m_{D^{*+}}-\frac{\vec{k}^2}{2m_{D^{*+}}}+i\frac{\Gamma_{D^{*+}}}{2}}\epsilon^{ijk}\epsilon^i(T_{cc}^{*+})p_3^j\epsilon^{kmn}p_1^m\epsilon^{n*}(\gamma),
\end{align}
and the other amplitude from the crossed-channel effects of the final-state identical $D^0$ particles is
\begin{align}
&i \mathcal{A}_{\text{a}}[T_{cc}^{*+} \to  \gamma(p_1)D^0(p_4)\pi^+(p_3)D^0(p_2)]=\frac{-i g_s \sqrt{2} \bar{g} \mu_{D^0}}{2\sqrt{2}F_{\pi}\sqrt{2m_{\pi^+}}}\frac{1}{t^0-m_{D^{*0}}-\frac{\vec{t}^2}{2m_{D^{*0}}}+i\frac{\Gamma_{D^{*0}}}{2}}\nonumber\\
&\, \times \frac{1}{l^0-m_{D^{*+}}-\frac{\vec{l}^2}{2m_{D^{*+}}}+i\frac{\Gamma_{D^{*+}}}{2}}\epsilon^{ijk}\epsilon^i(T_{cc}^{*+})p_3^j\epsilon^{kmn}p_1^m\epsilon^{n*}(\gamma),
\end{align}

The LO amplitudes from the $D^{*+}D^0/D^{*0}D^+$ rescattering diagrams including the crossed-channel contributions in Figs.~\ref{fig_TsDDpicgFDb} and \ref{fig_TsDDpicgFDc} are
\begin{align}
&i \mathcal{A}_{\text{b}}[T_{cc}^{*+} \to  \gamma(p_1)D^0(p_2)\pi^+(p_3)D^0(p_4)]=\frac{i g_s \sqrt{2} \bar{g} \mu_{D^0} C_{0D1}}{2\sqrt{2} F_{\pi} \sqrt{2m_{\pi^+}}}\frac{1}{(p_3^0+p_4^0)-m_{D^{*+}}-\frac{(\vec{p}_{3}+\vec{p}_{4})^2}{2m_{D^{*+}}}+i\frac{\Gamma_{D^{*+}}}{2}}\nonumber\\
&\, \times \epsilon^{ijk}\epsilon^i(T_{cc}^{*+}) p_3^j \epsilon^{kmn}p_1^m\epsilon^{n*}(\gamma)I(p_1),
\end{align}
\begin{align}
&i \mathcal{A}_{\text{b}}[T_{cc}^{*+} \to  \gamma(p_1)D^0(p_4)\pi^+(p_3)D^0(p_2)]=\frac{i g_s \sqrt{2} \bar{g} \mu_{D^0} C_{0D1}}{2\sqrt{2} F_{\pi} \sqrt{2m_{\pi^+}}}\frac{1}{(p_3^0+p_2^0)-m_{D^{*+}}-\frac{(\vec{p}_{3}+\vec{p}_{2})^2}{2m_{D^{*+}}}+i\frac{\Gamma_{D^{*+}}}{2}}\nonumber\\
&\, \times \epsilon^{ijk}\epsilon^i(T_{cc}^{*+}) p_3^j \epsilon^{kmn}p_1^m\epsilon^{n*}(\gamma)I(p_1),
\end{align}
\begin{align}
&i \mathcal{A}_{\text{c}}[T_{cc}^{*+} \to  \gamma(p_1)D^0(p_2)\pi^+(p_3)D^0(p_4)]=\frac{-i g_s \sqrt{2} \bar{g} \mu_{D^+} C_{0D1ex}}{2\sqrt{2} F_{\pi} \sqrt{2m_{\pi^+}}}\frac{1}{(p_3^0+p_4^0)-m_{D^{*+}}-\frac{(\vec{p}_{3}+\vec{p}_{4})^2}{2m_{D^{*+}}}+i\frac{\Gamma_{D^{*+}}}{2}}\nonumber\\
&\, \times \epsilon^{ijk}\epsilon^i(T_{cc}^{*+}) p_3^j \epsilon^{kmn}p_1^m\epsilon^{n*}(\gamma)I(p_1),
\end{align}
\begin{align}
&i \mathcal{A}_{\text{c}}[T_{cc}^{*+} \to  \gamma(p_1)D^0(p_4)\pi^+(p_3)D^0(p_2)]=\frac{-i g_s \sqrt{2} \bar{g} \mu_{D^+} C_{0D1ex}}{2\sqrt{2} F_{\pi} \sqrt{2m_{\pi^+}}}\frac{1}{(p_3^0+p_2^0)-m_{D^{*+}}-\frac{(\vec{p}_{3}+\vec{p}_{2})^2}{2m_{D^{*+}}}+i\frac{\Gamma_{D^{*+}}}{2}}\nonumber\\
&\, \times \epsilon^{ijk}\epsilon^i(T_{cc}^{*+}) p_3^j \epsilon^{kmn}p_1^m\epsilon^{n*}(\gamma)I(p_1),
\end{align}
where the masses $m_1$, $m_2$, and $m_3$ in the loop integrals $I(p_i)$ are taken to be the masses of $D^{*0}$, $D^{*+}$, and $D^0$ in Fig.~\ref{fig_TsDDpicgFDb}, and the masses of $D^{*+}$, $D^{*0}$, and $D^+$ in Fig.~\ref{fig_TsDDpicgFDc}, respectively.

The NLO amplitudes from the $D^*\pi$ rescattering diagrams considering the crossed-channel effects in Figs.~\ref{fig_TsDDpicgFDd} and \ref{fig_TsDDpicgFDe} are
\begin{align}
&i \mathcal{A}_{\text{d}}[T_{cc}^{*+} \to  \gamma(p_1)D^0(p_2)\pi^+(p_3)D^0(p_4)]=\frac{i g_s \sqrt{2} \bar{g} \mu_{D^0} C_{\pi3}}{4\sqrt{2} F_{\pi} m_{\pi^+}\sqrt{2m_{\pi^+}}}\frac{1}{(p_1^0+p_2^0)-m_{D^{*0}}-\frac{(\vec{p}_{1}+\vec{p}_{2})^2}{2m_{D^{*0}}}+i\frac{\Gamma_{D^{*0}}}{2}}\nonumber\\
&\, \times \epsilon^{ijk}\epsilon^i(T_{cc}^{*+}) p_4^j \epsilon^{kmn}p_1^m\epsilon^{n*}(\gamma)\left[I^{(1)}(p_4)-I(p_4)\right],
\end{align}
\begin{align}
&i \mathcal{A}_{\text{d}}[T_{cc}^{*+} \to  \gamma(p_1)D^0(p_4)\pi^+(p_3)D^0(p_2)]=\frac{i g_s \sqrt{2} \bar{g} \mu_{D^0} C_{\pi3}}{4\sqrt{2} F_{\pi} m_{\pi^+}\sqrt{2m_{\pi^+}}}\frac{1}{(p_1^0+p_4^0)-m_{D^{*0}}-\frac{(\vec{p}_{1}+\vec{p}_{4})^2}{2m_{D^{*0}}}+i\frac{\Gamma_{D^{*0}}}{2}}\nonumber\\
&\, \times \epsilon^{ijk}\epsilon^i(T_{cc}^{*+}) p_2^j \epsilon^{kmn}p_1^m\epsilon^{n*}(\gamma)\left[I^{(1)}(p_2)-I(p_2)\right],
\end{align}
\begin{align}
&i \mathcal{A}_{\text{e}}[T_{cc}^{*+} \to  \gamma(p_1)D^0(p_2)\pi^+(p_3)D^0(p_4)]=\frac{i g_s \bar{g} \mu_{D^0} C_{\pi3ex}}{4\sqrt{2} F_{\pi} \sqrt{m_{\pi^+}m_{\pi^0}}\sqrt{2m_{\pi^0}}}\frac{1}{(p_1^0+p_2^0)-m_{D^{*0}}-\frac{(\vec{p}_{1}+\vec{p}_{2})^2}{2m_{D^{*0}}}+i\frac{\Gamma_{D^{*0}}}{2}}\nonumber\\
&\, \times \epsilon^{ijk}\epsilon^i(T_{cc}^{*+}) p_4^j \epsilon^{kmn}p_1^m\epsilon^{n*}(\gamma)\left[I^{(1)}(p_4)+I(p_4)\right],
\end{align}
\begin{align}
&i \mathcal{A}_{\text{e}}[T_{cc}^{*+} \to  \gamma(p_1)D^0(p_4)\pi^+(p_3)D^0(p_2)]=\frac{i g_s \bar{g} \mu_{D^0} C_{\pi3ex}}{4\sqrt{2} F_{\pi} \sqrt{m_{\pi^+}m_{\pi^0}}\sqrt{2m_{\pi^0}}}\frac{1}{(p_1^0+p_4^0)-m_{D^{*0}}-\frac{(\vec{p}_{1}+\vec{p}_{4})^2}{2m_{D^{*0}}}+i\frac{\Gamma_{D^{*0}}}{2}}\nonumber\\
&\, \times \epsilon^{ijk}\epsilon^i(T_{cc}^{*+}) p_2^j \epsilon^{kmn}p_1^m\epsilon^{n*}(\gamma)\left[I^{(1)}(p_2)+I(p_2)\right],
\end{align}
where the masses $m_1$, $m_2$, and $m_3$ in the loop integrals $I^{(1)}(p_i)$ or $I(p_i)$ are taken to be the masses of $D^{*+}$, $D^{*0}$, and $\pi^+$ in Fig.~\ref{fig_TsDDpicgFDd}, and the masses of $D^{*0}$, $D^{*+}$, and $\pi^0$ in Fig.~\ref{fig_TsDDpicgFDe}, respectively.

\bibliography{TccStar_refs.bib}

\end{document}